\documentclass[a4paper,12pt]{article}

\pdfoutput=1 

\usepackage[hmargin=.7in,vmargin=1.1in]{geometry}
\usepackage{indentfirst}
\usepackage{amsfonts}
\usepackage{mathrsfs}
\usepackage{amsmath}
\usepackage{amssymb}
\usepackage{authblk}
\usepackage{cite}
\usepackage{xcolor}
\usepackage{mathtools}
\usepackage{tensor}
\usepackage{physics}
\usepackage{graphicx}
\usepackage{bm}
\usepackage{upgreek}
\usepackage{braket}
\usepackage{color,soul}
\usepackage{csquotes}
\usepackage{caption}
\usepackage{subcaption}

\usepackage[bookmarksnumbered=true,bookmarksopen=true]{hyperref}
 \hypersetup{colorlinks,
             linkcolor=[rgb]{0,0.3,0.6}, 
             citecolor=[rgb]{0,0.3,0.6}, 
             urlcolor=[rgb]{0,0.3,0.6}}

\newcommand\mi{\mathrm{i}}
\newcommand\me{\mathrm{e}}

\newcommand\pp{\uppi}
\newcommand{\dif}{\mathrm{d}}

\DeclareMathOperator{\diag}{diag}

\begin{document}

\title{\Large\textbf{Singularities of regular black holes and the monodromy method for asymptotic quasinormal modes}}

\author[a,b]{Chen Lan\thanks{lanchen@nankai.edu.cn}}
\author[c,d]{Yi-Fan Wang\thanks{yfwang@thp.uni-koeln.de}}
	
\affil[a]{\normalsize{\em Department of Physics, Yantai University, Yantai 264005, China}}	
\affil[b]{\normalsize{\em School of Physics, Nankai University, Tianjin 300071, China}}

\affil[c]{\normalsize{\em Niologic GmbH, Max-Ernst-Str.\ 4, 50354 Hürth, Germany}}
\affil[d]{\normalsize{\em Institute for Theoretical Physics, University of Cologne, Zülpicher Str.\ 77a, 50937 Cologne, Germany}}

\date{}

\maketitle

\begin{abstract}
We use the monodromy method to investigate the asymptotic quasinormal modes of regular black holes based on the explicit Stokes portraits.
We find that, for regular black holes with spherical symmetry and a single shape function, the analytical forms of the asymptotic frequency spectrum are not universal and do not depend on the multipole number but on the presence of complex singularities and the trajectory of asymptotic solutions along the Stokes lines.
\end{abstract}

\tableofcontents

\section{Introduction}
\label{sec:intr}

Regular black holes (RBHs) are a collection of black holes (BHs) that do not have
singularities in spacetime, especially at their centers \cite{Ayon-Beato:1998hmi,Hayward:2005gi,Ansoldi:2008jw}, i.e.,
all the curvature invariants of BHs, such as the Kretschmann scalar, are finite everywhere. 
This definition is related to Markov's limiting curvature conjecture \cite{Markov:1982lcc,Frolov:2016pav}, which states that the curvature invariants should be
restricted by a universal value. There is also a well-known definition of the regularity of spacetime through the completeness of null and timelike geodesics \cite{Hawking:1973uf,Wald:1984rg}. Unfortunately, these two definitions are not equivalent. Some BHs have finite curvature everywhere, but the spacetime contain  incomplete geodesics, e.g., the Taub--NUT BH \cite{Misner:1969sf,Kagramanova:2010bk}. Others are the opposite, namely, the geodesics in spacetime are complete everywhere, but the curvatures can be divergent in a certain area, e.g., a wormhole model in Ref.\ \cite{Olmo:2015bya}.    

The study of RBHs dates back to Sakharov and Gliner \cite{Sakharov:1966aja,Gliner:1966} and is interesting because RBHs are \emph{not} restricted by the Penrose singularity theorem \cite{Borde:1996df}. 
The spacetime singularity can be avoided by replacing the center with a de Sitter core\cite{Dymnikova:1992ux}.
The first RBH model was implemented by Bardeen \cite{Bardeen:1968non}, 
in which a RBH was obtained by a formal modification of the Schwarzschild BH. 
A significant breakthrough came in 1998, when Ayon-Beato and Garcia offered the first interpretation of matter, in terms of classical fields, that generates RBHs \cite{Ayon-Beato:1998hmi}. 
In their follow-up research \cite{Ayon-Beato:2000mjt}, 
Ayon-Beato and Garcia also showed that the Bardeen BH can be interpreted by the nonlinear magnetic monopole. 
This idea was generalized further by Fan and Wang \cite{Fan:2016hvf}. 
Now, {\em all} spherically symmetric RBHs with singular shape functions can be interpreted in the context of nonlinear electrodynamics, 
i.e., given the metric of a RBH, the Lagrangian of matter generation can be derived in terms of electrodynamic field strength.
These results stimulated further research on RBHs, including not only their construction \cite{Ayon-Beato:2004ywd,Hayward:2005gi,Nicolini:2008aj,Bambi:2013ufa,Balart:2014cga,Lan:2021ayk,Lan:2022bld}, 
but also the interpretation of RBH models \cite{Bonanno:2000ep,Bronnikov:2005gm,Modesto:2011kw,Modesto:2010uh}
and various physical phenomena in RBHs spacetime, such as BH thermodynamics \cite{Lan:2020wpv,Lan:2021ngq}, BH shadows \cite{Li:2013jra,Abdujabbarov:2016hnw},  and quasinormal modes (QNMs)\cite{Flachi:2012nv,Fernando:2012yw,Toshmatov:2015wga}.

The research program for RBHs may differ from that of traditional BHs with singularities.
In the latter, the full action of the system is first given,
and the BH solution can then be solved using the Euler--Lagrange equation obtained from
the variational principle.
In the former, the metric of a RBH can be constructed from the aspect of finite curvature invariants.
Subsequently, information on the matter source can be deduced within a specific context that facilitates the solution, such as nonlinear electrodynamics \cite{AyonBeato:2000zs,Bronnikov:2000vy,Bronnikov:2006fu,Zhang:2016ilt}, in the presence of a phantom scalar field \cite{Bronnikov:2005gm,Bronnikov:2012ch} or modified gravity\cite{Berej:2006cc,Moffat:2014aja}.

The area/entropy spectrum of BHs is known to be related to their QNMs \cite{Hod:1998vk,Bronnikov:2012ch}
and, specifically, to the imaginary component of QNMs in the large damping limit, i.e.\ {\em asymptotic} QNMs (AQNMs) \cite{Maggiore:2007nq}.
In other words, 
the area/entropy spectrum is obtained via 
the
Bohr--Sommerfeld quantization of an adiabatic invariant \cite{Kunstatter:2002pj}
 constructed using the AQNMs, 
whereas
the AQNMs of BHs can be determined by the so-called monodromy approach \cite{Motl:2003cd},
in which singularities and Stokes lines play crucial roles \cite{Natario:2004jd}. 
As a result, studying the quantum entropy spectrum of RBHs requires a thorough understanding of AQNMs. 
The goal of this study is to investigate the AQNMs of RBHs in greater detail using a clear Stokes diagram.

For a traditional BH with a singularity at the center, such as 
the
Schwarzschild BH, 
the Stokes lines emit from the central singularity.
Thus, when the asymptotic solutions of the Regge--Wheeler master equation
rotate from one Stokes line to another around the zero point of the radial coordinate (i.e., the center of the BH), 
the change is not trivial,
and the monodromy along a closed Stokes line will give an analytical expression for AQNMs, 
see e.g.\ \cite{Berti:2009kk,Konoplya:2011qq}.

For a RBH \cite{Ansoldi:2008jw},
the zero point of the metric is no longer singular in most cases, 
and the corresponding Stokes lines do not converge to that point, 
which implies that the rotation of the asymptotic solution around the zero point is trivial.
Several attempts have been made by ignoring this problem \cite{Flachi:2012nv,Giri:2006rc},
where the Stokes lines of RBHs are borrowed from Reissner--Nordstr\"om (RN)
or RN-like BHs
based on the similarity of the metrics at infinity, and the asymptotic frequencies of these models have a universal form, see Eq.(12) in \cite{Flachi:2012nv}.
In other words, to obtain the monodromy, the RBH is regarded as a singular BH.
Therefore, one may ask 
what the {\em true} Stokes lines of RBHs are
and what the monodromy based on these is.
In the current study, we give a detailed discussion.

This paper is organized as follows: 
In Sec.\ \ref{sec:singularity},
we illustrate and classify the singularities of RBHs from the perspective of complex analysis, 
which can be regarded as a prelude for the calculation of AQNMs in the subsequent sections.

After introducing an alternative approach to displaying the Stokes diagram in Sec.\ \ref{sec:QNM-scalar},
 we calculate the AQNMs of RBHs that still have complex singularities as the poles of the shape functions based on the explicit Stokes portrait in Sec.\ \ref{sec:poles}.
Sec.\ \ref{sec:other} contains other examples in which the RBHs have essential singularities or no singularities at all.
 
We also provide a brief comparison of the monodromy method and complex WKB method from the perspective of the Stokes portrait in 
Sec.\ 
\ref{sec:comparison},
and a conclusion is given in Sec.\ \ref{sec:concl}.

\section{Singularities of regular black holes}
\label{sec:singularity}

Let us start with spherically symmetric BHs with 
a
single {\em shape function} $f(r)$ in the metric 
 \begin{equation}
 \label{eq:metric}
     g_{\mu\nu}=\diag\{-f(r), f^{-1}(r), r^2, r^2\sin^2(\theta)\},
 \end{equation} 
where we suppose that $f(r)$ is of the following form: 
\begin{equation}
\label{eq:shape-function}
f(r)=1-\frac{2M}{r} \sigma\left(r\right).
\end{equation}
Here, $M$ is the BH mass, and $\sigma(r)$ is a dimensionless function of the radial coordinate.
A BH with the metric in Eq.\ \eqref{eq:metric} is regular%
\footnote{The most satisfactory definition of spacetime being regular is 
formulated via geodesic completeness \cite{Hawking:1973uf,Wald:1984rg}, 
i.e., a spacetime is regular if it does not possess at least any incompletely null and timelike geodesics.
Nevertheless, not a few examples can be given of the failure of this definition, e.g. Ref.\ \cite{Geroch:1968ut}.
The ``regularity'' applied in the current paper refers to the finiteness of curvature invariants, 
which is not equivalent to geodesic completeness but has an intersection with it. 
This definition is widely used in the field of RBHs, although it is flawed in some models, e.g.\ Taub--NUT spacetime \cite{Kagramanova:2010bk}.}
if all its curvature invariants referring to the Riemann tensor are finite for $r\in [0,\infty)$,
e.g.\ 
the Ricci curvature $R\coloneqq g^{\mu\nu} R_{\mu\nu}$ and
Weyl curvature $W\coloneqq W_{\mu\nu\alpha\beta}W^{\mu\nu\alpha\beta}$, 
where $R_{\mu\nu}$ and $W_{\mu\nu\alpha\beta}$ are the Ricci and Weyl tensors respectively.
From a physical perspective, 
``regularity'' implies that there are no gravitational singularities within the BH horizons. Furthermore, {\em all} RBHs with the Eq.\ \eqref{eq:metric} metric can be interpreted as magnetic monopoles \cite{Fan:2016hvf}.

For the Eq.\ \eqref{eq:metric} metric, we can compute the Ricci and Weyl curvatures
\begin{equation}
\label{eq:curvature}
R=\frac{2 M }{r^2} \left(2 \sigma '+r \sigma ''\right),\quad
W=\frac{4 M^2}{3 r^6}  \left[r \left(r \sigma ''-4 \sigma '\right)+6 \sigma\right]^2,
\end{equation}
from which we note that the singularities of the curvature invariants are unavoidable if the function $f(r)$ has singularities in $r\in [0,\infty)$. 
Additional restricted conditions for $f(r)$ that guarantee the finiteness of the curvature invariants at $r=0$
have been discussed, for example, in Ref. \cite{Balart:2014cga,Frolov:2016pav,Fan:2016hvf,Lan:2021ngq}.

Furthermore, the Ricci and Weyl curvatures can be simplified if we replace
$\sigma\to s_1(r)/(2 M r)$ for $R$ and $\sigma \to \sqrt{3} r^2 s_2(r)/(2M)$ for $W$,
yielding
\begin{equation}
\label{eq:weyl}
R=\frac{s_1''(r)}{r^2},
\qquad
W=r^2 s_2''(r)^2,
\end{equation}
which provides convenience when constructing RBHs with various types of {\em mathematical} singularities.

In contrast, according to the monodromy method, as we compute the AQNMs for BHs with gravitational singularities at $r=0$, 
such as Schwarzschild and RN BHs,
the radial coordinate $r$ in the Regge–Wheeler master
equation is analytically continued into the complex plane, and 
the {\em physical} singularities (gravitational singularities and BH horizons)\footnote{In contrast to the \emph{mathematical} singularities in the analytically continued complex plane that will be illustrated below.} play a key role \cite{Motl:2003cd}. 
Therefore, it is natural to wonder what to do when dealing with the AQNMs of RBHs if there are no gravitational singularities.

In fact, the mechanisms for removing the singularities of BHs are considerbaly different from each other
from the perspective of complex analysis, and these differences are reflected in the calculations of AQNMs.
Therefore, we first divide  
the RBHs into three classes according to the types of singularities of the curvature invariants after analytically continuing the radial coordinate into the complex plane:
\begin{enumerate}
\item The curvature invariants are
entire functions of $r$, i.e., they have either
no singularities on the entire complex plane $\mathbb{C}$ or removable singularities at $r=0$. 
\item The curvature invariants are meromorphic functions on $\mathbb{C}$. In other words, the point $r=0$ is a \emph{regular} point;
however, the singularities (as poles) still exist and are located in the ranges beyond the non-negative axis of $r$, i.e., the singularities of the curvature invariants are dragged into the non-physical domain.
\item  $r=0$ is an \emph{essential} singularity of the curvature invariants. Via Picard's great theorem \cite{shabat2004intro},  RBHs with an essential singularity at $r=0$ must have infinitely many {\em complex} horizons on any punctured neighborhood of $r=0$.
\end{enumerate}

Several BH examples can be used to help understand this classification.
First, we propose a RBH model with\footnote{We will not focus on the field interpretation of the metric here because our goal is to demonstrate the classification of RBHs' singularities, which display their influence in the computations of AQNMs in the following. In reality, because every RBH metric can have a nonlinear electrodynamic interpretation, determining the action functional of the matter field in nonlinear electrodynamics is not a challenging task.}
\begin{equation}
\label{eq:bounce}
\sigma = 
\frac{M r}{Q^2}
-\frac{Q^2}{16 M r}
\sin^2\left(\frac{4 M r}{Q^2}\right),
\end{equation}
where $Q$ is a parameter 
introduced to balance the dimensions.

To reduce the parameters, we rescale 
$r\to   l^{2/3} (2M)^{1/3} z$ and $Q\to 2^{7/6} l^{1/3} M^{2/3}$,
where $z$ is a dimensionless 
radial coordinate, and $l$ is a parameter of length dimension and is expected to be 
the Planck length or of the same order.
Then, the horizons $z_{\rm H}$ can be determined using the equation
\begin{equation}
\label{eq:def-lambda}
\lambda\coloneqq\left(\frac{l}{2 M}\right)^{2/3}=\frac{1}{4}-\frac{\sin ^2(z_{\rm H})}{4 z_{\rm H}^2}.
\end{equation}
Because sine is a transcendental function, there must be infinitely many {\em complex} horizons for any $\lambda\in\mathbb{R}$ and no Picard exceptional values according to 
Picard's little theorem \cite{shabat2004intro,Goldberg2008vdm}; however, the number of complex horizons is finite in any circle around $z=0$.

We also note that the number of {\em physical} horizons increases discretely with increasing $\lambda$ in the interval $(0.238, 1/4]$. 
The model has only one horizon when $\lambda\lesssim 0.238$,  two horizons when 
$\lambda\approx 0.238$, and an infinite number of horizons when $\lambda=1/4$.
The Ricci and Weyl curvatures can be computed as
\begin{equation}
R=\frac{\sin^2 (z)}{l^2 z^2},\qquad
W=\frac{\sin ^2(z) }{3 l^4 z^8}\left[\left(z^2-3\right) \sin (z)+3 z \cos (z)\right]^2,
\end{equation}
where $z=0$ is a removable singularity  belonging to the first type above. 
Moreover, it is easy to verify that this model meets the dominant energy condition (DEC) \cite{Curiel:2014zba}.

To understand another case in the first type, 
we propose a model of a RBH with  
\begin{equation}
\label{eq:infinity}
\sigma=\me^{-\frac{r}{l}} \left(\frac{6 l}{r}+\frac{r}{l}+4\right)-\frac{6 l}{r}+2.
\end{equation}
Under the rescalings $r\to l^{2/3} (2 M)^{1/3} z, l \to l^{2/3} (2 M)^{1/3}$, we can fix the horizons via
\begin{equation}
\label{eq:lambda-2}
\lambda=\me^{-z_{\rm H}} \left(\frac{6}{z_{\rm H}^2}+\frac{4}{z_{\rm H}}+1\right)-\frac{6}{z_{\rm H}^2}+\frac{2}{z_{\rm H}},
\end{equation}
where $\lambda$ is formally the same as defined in Eq.\ \eqref{eq:def-lambda}.
When $\lambda\lesssim 0.173$, there are two {\em physical} horizons, 
whereas there are infinitely many {\em complex} horizons.
In addition, the DEC of this model also holds.
The typical curvature invariants are then
\begin{equation}
R=\frac{\me^{-z}}{l^2},\qquad
W=\frac{\me^{-2 z} }{3 l^4 z^8} 
\left[12 \me^z (z-6)
+z^4+6 z^3+24 z^2+60 z
+72\right]^2,
\end{equation}
which are surely finite in the physical domain. 
In contrast, as we analytically continue $z$ into the entire complex plane, 
although there are no singularities in the finite range 
divergence occures at complex infinity owing to the Stokes phenomenon of the function $\me^{-z}$.
In other words, as $z$ approaches infinity along a path restricted in the sector
\begin{equation}
\frac{\pp }{2}<\arg(z) <\frac{3 \pp }{2},
\end{equation}
the curvature invariants become divergent, namely
 complex infinity is an essential singularity of the model.

Now, we turn to the Hayward BH \cite{Hayward:2005gi} with $\sigma=r^3/\left(r^3+2Ml^2\right)$, 
where $l/(2M)\ll 1$.
The curvature invariants of this model are finite in the domain $r\in [0,\infty)$.
However, after analytically continuing $r$ into the complex plane, the Weyl curvature 
\begin{equation}
W= \frac{48 M^2 r^6 \left(r^3-4 l^2 M\right)^2}{\left(2 l^2 M+r^3\right)^6}
\end{equation}
starts having three singularities in the non-physical domain, 
which are determined using the algebraic equation $2 l^2 M+r^3=0$.
More precisely, the Hayward BH has three poles after the analytical continuation. 
This is what we have in the second type, i.e., the singularities are hidden in the non-physical domain.

Furthermore, we can determine that all three singularities are located in the outermost horizons. The proof is a typical application of Rouch\'e's theorem \cite{shabat2004intro}. 
First, we note from $2 l^2 M+r^3=0$ that all the singularities are located on a circle with the radius $\abs{r}= (2M l^2)^{1/3}$. Second, we can rewrite $f(r)=0$ as $r^3-2M r^2+2M l^2=0$, i.e., 
we obtain three complex horizons using the fundamental theorem of algebra. 
Third, on the circle $\abs{r}= (2M l^2)^{1/3}+\epsilon$, where $\epsilon\ll(2M l^2)^{1/3}$, we have $\abs{2M r^2}>\abs{r^3}+\abs{2M l^2}$ if $l/2M<1/(2\sqrt{2})\approx 0.35$ and $\epsilon/(2M) < \left(\sqrt{5}-1\right)/4\approx 0.31$. Finally, because $l/(2M)\ll 1$, we can conclude that there are two horizons 
in the circle $\abs{r}= (2M l^2)^{1/3}+\epsilon$ by applying Rouch\'e's theorem.
In other words, the three complex horizons are sandwiched between the outermost and next to outermost horizons.

At last, the widely discussed model \cite{Balart:2014cga} with $\sigma(r)= \exp[-P^2/(2Mr)]$  belongs to the third type.
After rescalings $r\to l^{2/3} \sqrt[3]{2 M} z,P\to \sqrt[3]{l} (2 M)^{2/3}$,
its curvature invariants can be represented by
\begin{equation}
R=\frac{\me^{-1/z}}{l^2 z^5},\qquad
W=\frac{\me^{-2/z}}{3 l^4 z^{10}} [6 (z-1) z+1]^2
\end{equation}
where $z=0$ becomes an essential singularity after analytical continuation into the complex plane.
A detailed discussion on this model is given in Sec.\ \ref{subsec:essential}.

We end this section with a comment on constructing {\em physical} RBHs, 
Eqs.\ \eqref{eq:bounce} and \eqref{eq:infinity}.
Here, ``physical'' implies that the RBHs must obey the DEC  \cite{Curiel:2014zba}.
Instead of the traditional method of selecting the well-behaved $\sigma$ \cite{Bronnikov:2006fu,Frolov:2016pav},
we start with the curvature invariants Eqs.\ \eqref{eq:curvature} and \eqref{eq:weyl} directly, 
and choose {\em strictly positive} and {\em finite} functions as curvatures,
solving $\sigma$-functions inversely from the differential equations.
The ``strict positivity'' of Ricci curvature is a minimum requirement  for satisfying the DEC, 
whereas the ``finiteness'' is the regularity in $r\in[0,\infty)$ and asymptotic flatness at infinity in their given sense.
This process apparently provides an efficient approach to establishing physical RBHs,
and it has indeed helped us create a series of models. 

However, several problems in the process must to be treated carefully.
For example, the existence of horizons and the integrality of $\sigma$.
In other words, the $\sigma$-functions obtained from this method do not always provide horizons for the models.
Because these problems go beyond the scope of the current study,
we leave them for future research.

Furthermore, in this study, we concentrate on not only {\em physical} RBHs, 
but also non-physical models within the framework of Einstein's gravity
to focus on how the singularities of RBHs affect the spectrum of ANMQs.

\section{Asymptotic 
quasinormal modes of regular black holes}
\label{sec:QNM-scalar}

\subsection{Perturbation of a massless scalar field}

The perturbation of a BH can be realized by laying a probing field onto the BH spacetime \cite{Konoplya:2011qq}.
For a massless scalar field without backreaction,
the perturbation reduces to the propagation equation of a scalar field,
\begin{equation}
\label{eq:KG-eq}
\frac{1}{\sqrt{-g}} \partial_\nu\left(
g^{\mu\nu} \sqrt{-g} \partial_\mu \Psi
\right)=0.
\end{equation}
In the spherically symmetric spacetime given by Eq.\ \eqref{eq:metric},
we can substitute the ansatz 
$\Psi = \me^{-\mi\omega t} Y_l(\theta,\phi) r^{-1} \psi(r)$
and then select the radial component
\begin{equation}
f\partial_r\left(f \partial_r \psi\right)
+\left(\omega^2- V\right)\psi=0,\quad
V=f\left[\frac{l(l+1)}{r^2}+\frac{f'}{r}\right],
\end{equation}
which is a second order differential equation. Or in the canonical form,
we have
\begin{equation}
\label{eq:master}
\psi ''(r)+
p\psi'+q \psi =0,   
\end{equation}
where
\begin{equation}
\label{eq:coefficients}
    p= \frac{f'}{f},\qquad
    q=\frac{\omega ^2-V}{f^2}.
\end{equation}
Then, substituting the shape function, Eq.\ \eqref{eq:shape-function},
 the effective potential $V$ can be represented in terms of $\sigma$ and its derivatives
\begin{equation}
\label{eq:potential}
V=\frac{l (l+1)}{r^2}+V_{\rm reg},
\end{equation}
where 
\begin{equation}
V_{\rm reg}=-2 l (l+1) M\frac{ \sigma(r)}{r^3}
+2 M \frac{\sigma(r)}{r^2}-2 M \frac{\sigma'(r)}{r},
\end{equation}
is a regular function at $r=0$ when $r\in[0,\infty)$,
because for RBHs, $\sigma\sim O(r^3)$ as $r\to 0$ \cite{Balart:2014cga,Frolov:2016pav,Fan:2016hvf,Lan:2021ngq}.
Thus, for any BHs without degenerate horizons (i.e., the horizons are the simple roots of the algebraic equation $f(r)=0$), one may find that
$r=0$ and horizons are regular singular points of the master equation Eq.\ \eqref{eq:master}.

Furthermore, 
to diagonalize
Eq.\ \eqref{eq:master},
we can define the tortoise coordinate by an indefinite integral,
\begin{equation}
\label{eq:tortoise-coord0}
    z=\int \frac{\dif r}{f(r)},
\end{equation}
or the solution of 
the differential equation
\begin{equation}
\label{eq:tortoise-coord}
\frac{\dif z}{\dif r}= \frac{1}{f(r)}.
\end{equation}
The master equation, Eq.\ \eqref{eq:master}, then becomes
\begin{equation}
\label{eq:master-tot}
\left\{  \frac{\dif{}^2 }{\dif z^2}+\omega^2
-V[r(z)]\right\}\psi(z)=0,
\end{equation}
The QNMs are the solutions of Eq.\ \eqref{eq:master-tot} under the following boundary conditions:
\begin{equation}
\begin{split}
    \psi&\sim \me^{-\mi \omega z},\quad z\to-\infty;\\
    \psi&\sim  \me^{\mi \omega z},\quad z\to \infty .
\end{split}
\end{equation}
Thus, the high damping limit of QNMs will be defined as AQNMs, i.e., $\abs{\Im(\omega)}\gg \abs{\Re(\omega)}$ as the overtone number approaches infinity for BHs with asymptotic flatness, see e.g., Refs.\ \cite{Berti:2009kk,Konoplya:2011qq}.

Our goal in the current section is to find the 
AQNMs of Eq.\ \eqref{eq:master-tot} using an asymptotic analysis known as the monodromy method, in which the Stokes lines play a pivotal role.
Therefore, before studying the AQNMs of specific models, let us take a close look at the geometry of Stokes lines. A detailed illustration of the monodromy method can be found in Refs.\ \cite{Motl:2003cd,Natario:2004jd,Moura:2021nuh}.

\subsection{Stokes portraits for black holes}
\label{ssec:stokes-lines}

To understand the geometric aspects of Stokes lines, 
we first extend the radial coordinate $r$
into the complex plane and explicitly rewrite it with its real and imaginary components $x$ and $y$.
We can then recast Eq.\ \eqref{eq:tortoise-coord0} as
\begin{equation}
z(x, y)\coloneqq \int^{x+\mi y} \frac{\dif r}{f(r)},
\end{equation}
where the tortoise coordinate $z$ is also complex in general.
Based on this representation, 
the real and imaginary parts of $z(x, y)$ can be regarded as two surfaces,
and $z(x, y)=C$ are two groups of contours on each surfaces, i.e.
\begin{equation}
\Re[z(x, y)]= c_1,\qquad 
\Im[z(x, y)]= c_2,\qquad
C=c_1+\mi c_2.
\end{equation}
The Stokes lines are defined as the zero-contours, $\Re[z(x, y)]=0$, on the first surface \cite{Motl:2003cd}.
The Stokes lines must be symmetric with respect to the $y$-axis, i.e.
\begin{equation}
\Re[z(x, y)]=\frac{1}{2}\left[z(x, y)+z(x, -y)\right]=\Re[z(x, -y)].
\end{equation}
The one-form $\dif \Re[z]$ corresponds to a vector field, 
\begin{equation}
\dif \Re[z]=  \Re\left[1/f(x+\mi y)\right] \dif x
-\Im\left[1/f(x+\mi y)\right] \dif y,
\end{equation}
whereas its Hodge dual (with respect to the Euclidean metric) or the \emph{Stokes field} \cite{white2010asymptotic}
\begin{equation}
\label{eq:stokes-field}
\star\!\dif \Re[z]= \Im\left[1/f(x+\mi y)\right]   \dif x+ \Re\left[1/f(x+\mi y)\right]\dif y
\end{equation}
is tangent to the contours at each point.
In the following, we also use the component notation to represent the Stokes field, i.e.
\begin{equation}
{\bm V}_{\rm st}=\left\{
\Im\left[1/f(x+\mi y)\right],\;
\Re\left[1/f(x+\mi y)\right]
\right\}.
\end{equation}
Moreover, it is easy to verify that both
the divergence and curl of the Stokes fields are constantly zero,
i.e. $\star\dif\star (\star\dif \Re[z])=0=\star\dif\,(\star\dif \Re[z])$.
In other words, the Stokes field $\star\dif \Re[z]$ is a harmonic one-form \cite{Nakahara:2003nw} because $z(r)$ is a holomorphic function, and its real and imaginary parts satisfy the Cauchy--Riemann equations.

The {\em critical points} of the Stokes field are defined as a set of points $\{(x,y)|\star\dif \Re[z(x,y)]=0\}$,
i.e., the zeros of $1/f(r)=0$ on the complex plane of $r\in\mathbb{C}$.
Meanwhile, the critical points of the field are also the singular points \cite{Shafarevich2013} of the Stokes lines.
Then, one can immediately deduce a fact about the point $r=0$: the origin  $r=0$ is a critical point for singular BHs but not for RBHs,
which implies that the Stokes lines of singular BHs converge to (or emit
from) zero, 
whereas for RBHs, the convergence points may not  include $r=0$.

\begin{figure}[!ht]
     \centering
     \begin{subfigure}[b]{0.445\textwidth}
         \centering
         \includegraphics[width=\textwidth]{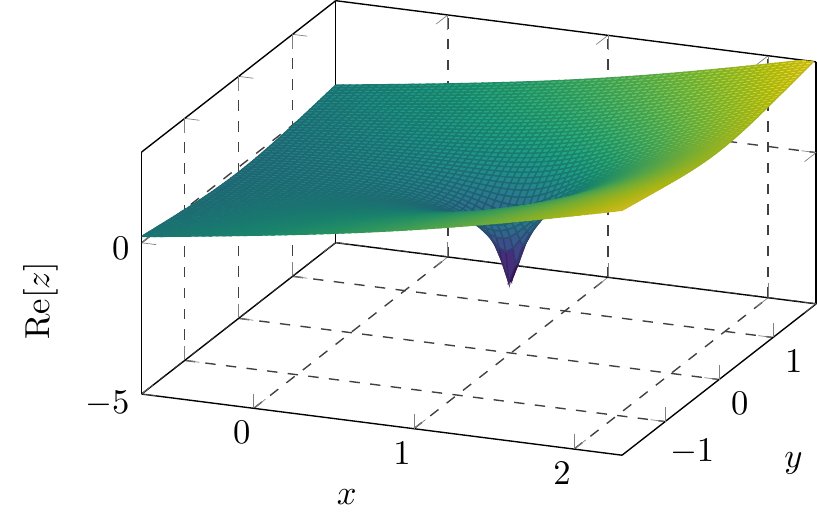}
         \caption{Schwarzschild BH}
         \label{fig:stokes-surface-schw}
     \end{subfigure}
     \begin{subfigure}[b]{0.455\textwidth}
         \centering
         \includegraphics[width=\textwidth]{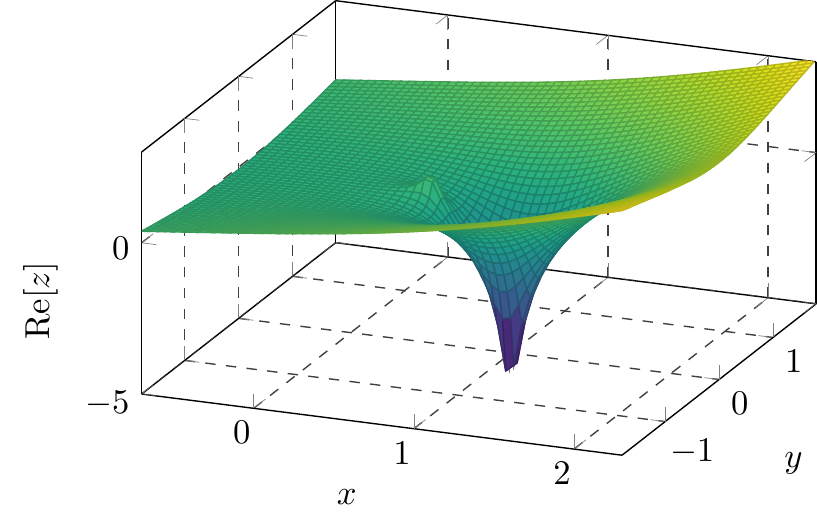}
         \caption{Reissner--Nordstr\"om BH}
         \label{fig:stokes-surface-rn}
     \end{subfigure}
      \captionsetup{width=.9\textwidth}
       \caption{Stokes surfaces $\Re[z(x, y)]$.}
            \label{fig:stokes-surface}
\end{figure}

\begin{figure}[!ht]
     \centering
     \begin{subfigure}[b]{0.45\textwidth}
         \centering
         \includegraphics[width=\textwidth]{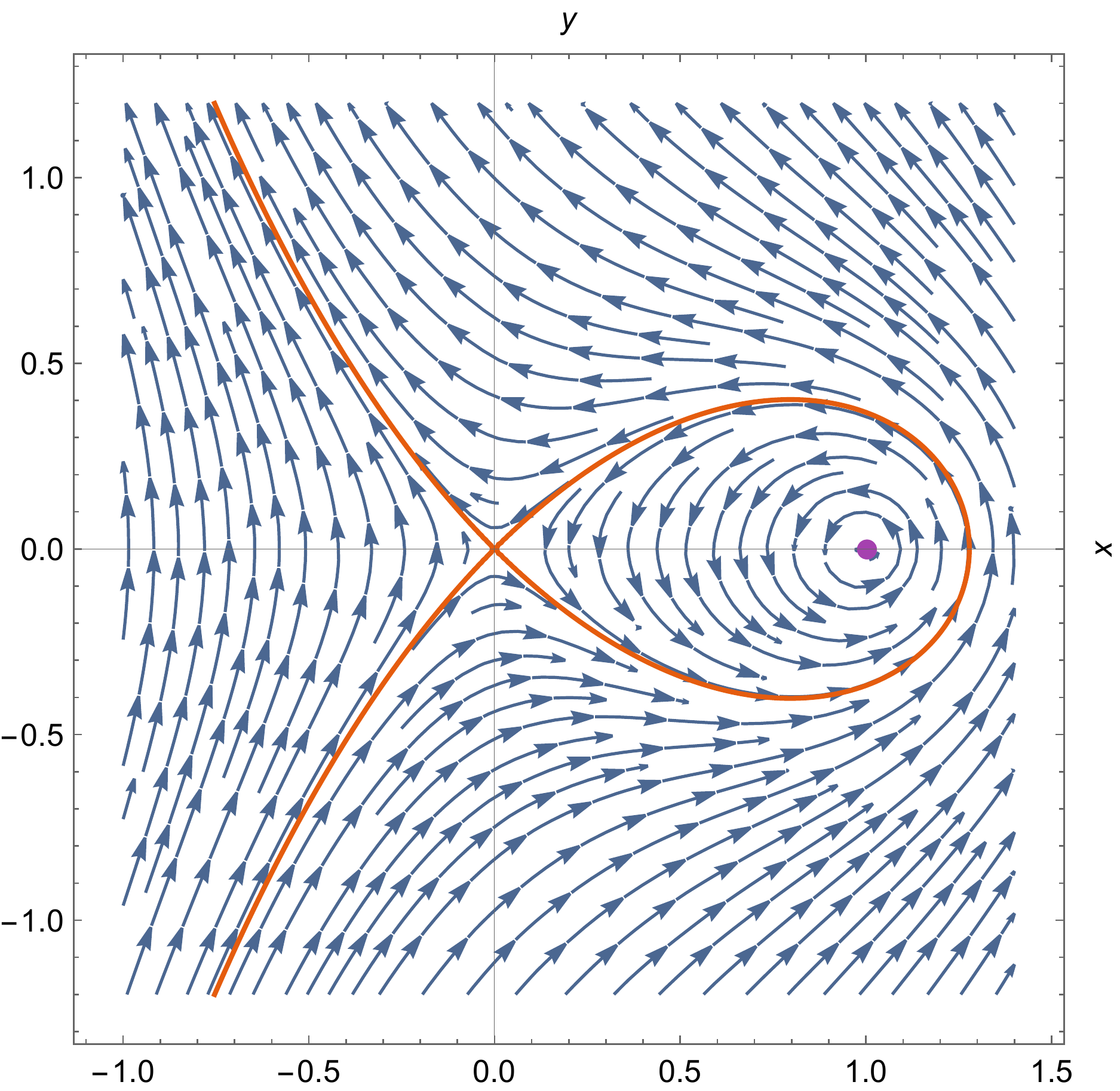}
         \caption{Schwarzschild BH}
         \label{fig:stokes-diag-schw}
     \end{subfigure}
     \begin{subfigure}[b]{0.45\textwidth}
         \centering
         \includegraphics[width=\textwidth]{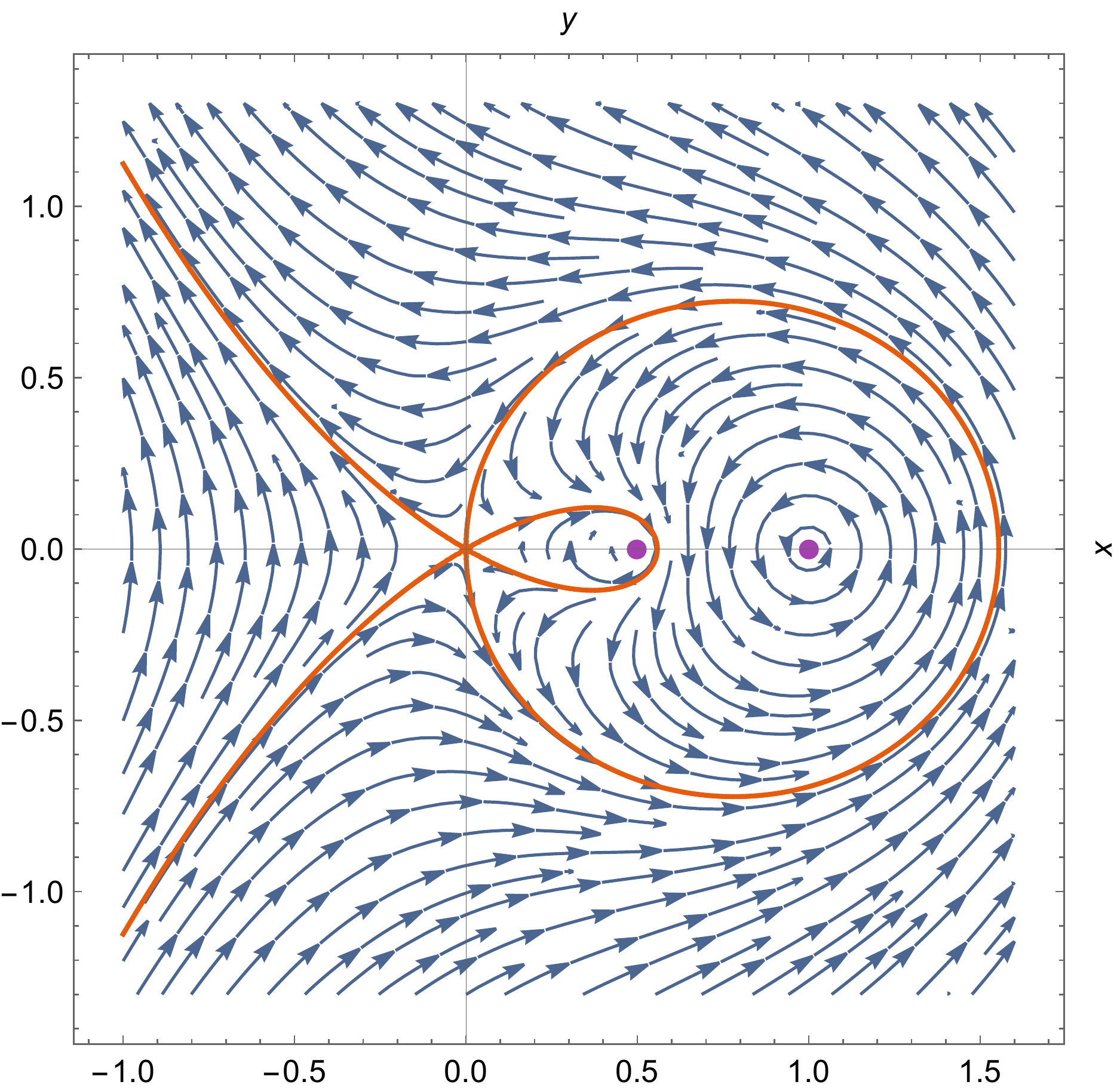}
         \caption{Reissner--Nordstr\"om BH}
         \label{fig:stokes-diag-rn}
     \end{subfigure}
      \captionsetup{width=.9\textwidth}
       \caption{Stokes diagrams in Cartesian coordinates; 
    $\star\dif\Re[z(x, y)]$--blue vector fields, and 
    the yellow-brown stream corresponds to the Stokes curves $\Re[z(x, y)]=0$, 
    where we set $k=1/2$ for RN BHs.}
         \label{fig:stokes-diag}
\end{figure}

Let us first review two integrable examples\footnote{The integrability refers to the differential equation Eq.\ \eqref{eq:tortoise-coord}.}, Schwarzschild and RN BHs.
The two plots in Fig.\ \ref{fig:stokes-surface} show the \emph{Stokes surfaces} $\Re[z(x, y)]$,
where the peaks on the surfaces correspond to the horizons.
Fig.\ \ref{fig:stokes-diag} is a \emph{Stokes portrait/diagram}, 
which consists of the
vector field $\star\dif \Re[z]$ (blue), Stokes lines $\Re[z]$=0 (yellow-brown) and complex horizons (purple). 
The Stokes lines can then be understood as the integral curves passing through the critical points 
in the Stokes portrait. Meanwhile, the complex horizons must be located inside the closed parts of the Stokes lines.

For the Schwarzschild BH with the shape function $f=1-1/r$,
the Stokes vector field reads
\begin{equation}
{\bm V}_{\rm st}=\left\{\frac{-y}{(x-1)^2+y^2},1+\frac{x-1}{(x-1)^2+y^2}\right\}.
\end{equation}
The origin $r=(0,0)$ on the complex plane is a critical point, i.e.,
this Stokes line has a self-intersection at the origin.
Near this point $(0,0)$, there is an approximation, ${\bm V}_{\rm st}\sim\left\{-y, -x\right\}$.
The Stokes line of the Schwarzschild BH can be integrated out analytically,
\begin{equation}
\label{eq:stokes-schwarzchild}
(x-1)^2+y^2=\me^{-2 x},
\end{equation}
which is a transcendental curve.
Moreover, around the origin $r=(0,0)$, Eq.\ \eqref{eq:stokes-schwarzchild} reduces to $y=\pm x$, i.e., 
there are $2\times 2=4$ lines emitting from the origin, 
and the angle between any two adjacent lines is $\pp/2$.

For the RN BH with the shape function $f=(r-1)(r-k)/r^2$,
where $1\ge k >0$, the Stokes field is
\begin{equation}
\begin{split}
{\bm V}_{\rm st}=\Big\{
&-\frac{x y [k (x-2)+x]+(k+1) y^3}
{\left[(x-1)^2+y^2\right] \left[(x-k)^2+y^2\right]}
,\\
&\frac{-y^2 \left(k x+k-2 x^2+x\right)-(x-1) x^2 (k-x)+y^4}
{\left[(x-1)^2+y^2\right] \left[(x-k)^2+y^2\right]}
\Big\}.
\end{split}
\end{equation}
The origin $r=(0, 0)$ is also a self-intersecting point,
near which there is an asymptotic behavior,
 ${\bm V}_{\rm st}\sim\left\{2 x y/k, (x^2-y^2)/k\right\}$.
The Stokes line for this case is also a transcendental curve,
\begin{equation}
(x-1)^2+y^2=\me^{2 (k-1) x}  k^{-2 k^2} \left[(k-x)^2+y^2\right]^{k^2}.
\end{equation}
Around the origin $r=(0, 0)$,  $x=0$ and $x=\pm \sqrt{3} \,y$, 
i.e., there are $3\times 2=6$ lines  emitting from the origin,
and the angle between any two adjacent lines is $\pp/3$.

\begin{figure}[!ht]
     \centering
     \begin{subfigure}[b]{0.445\textwidth}
         \centering
         \includegraphics[width=\textwidth]{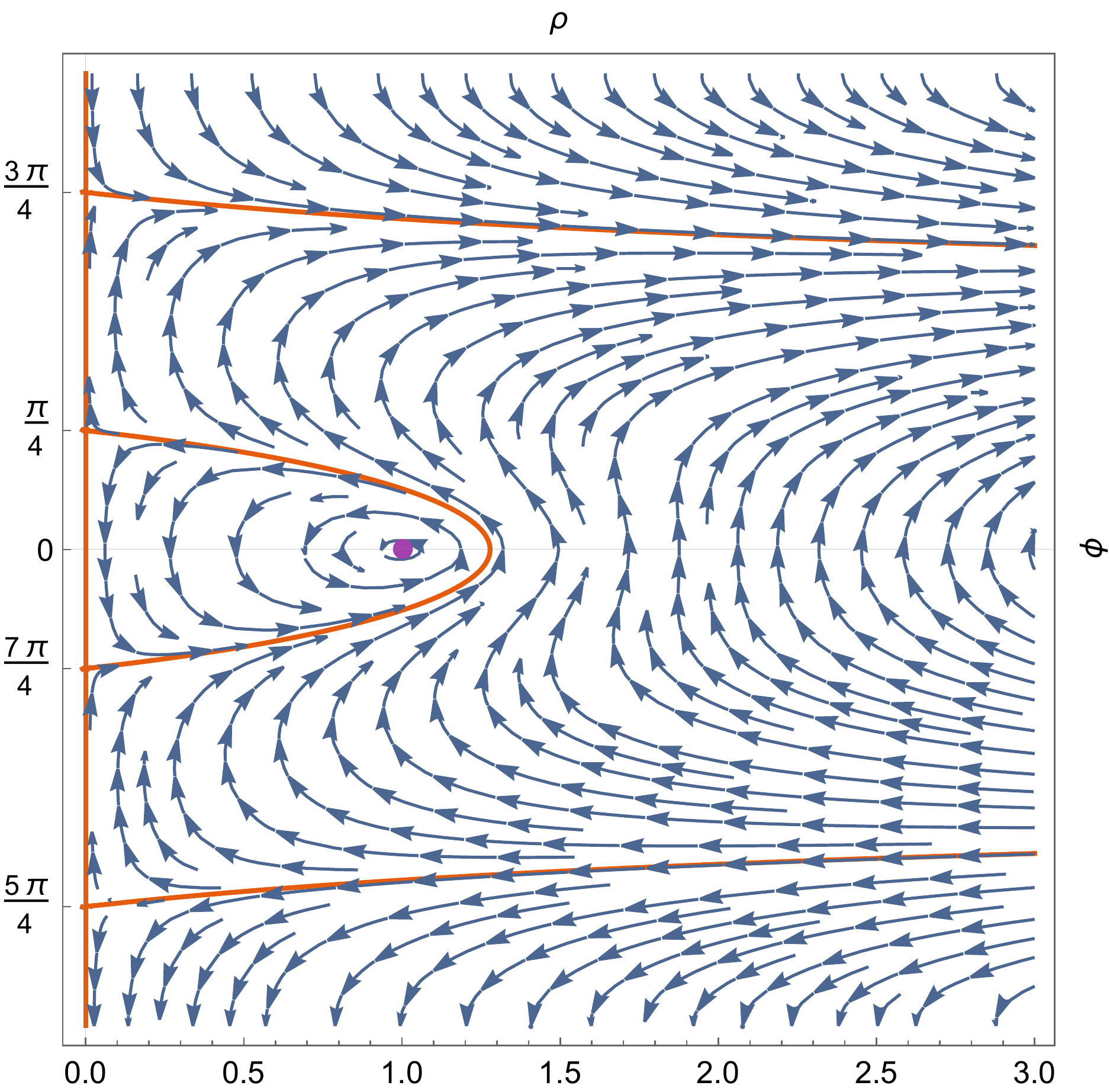}
         \caption{Schwarzschild BH}
         \label{fig:stokes-diag-polar-sch}
     \end{subfigure}
     \begin{subfigure}[b]{0.455\textwidth}
         \centering
         \includegraphics[width=\textwidth]{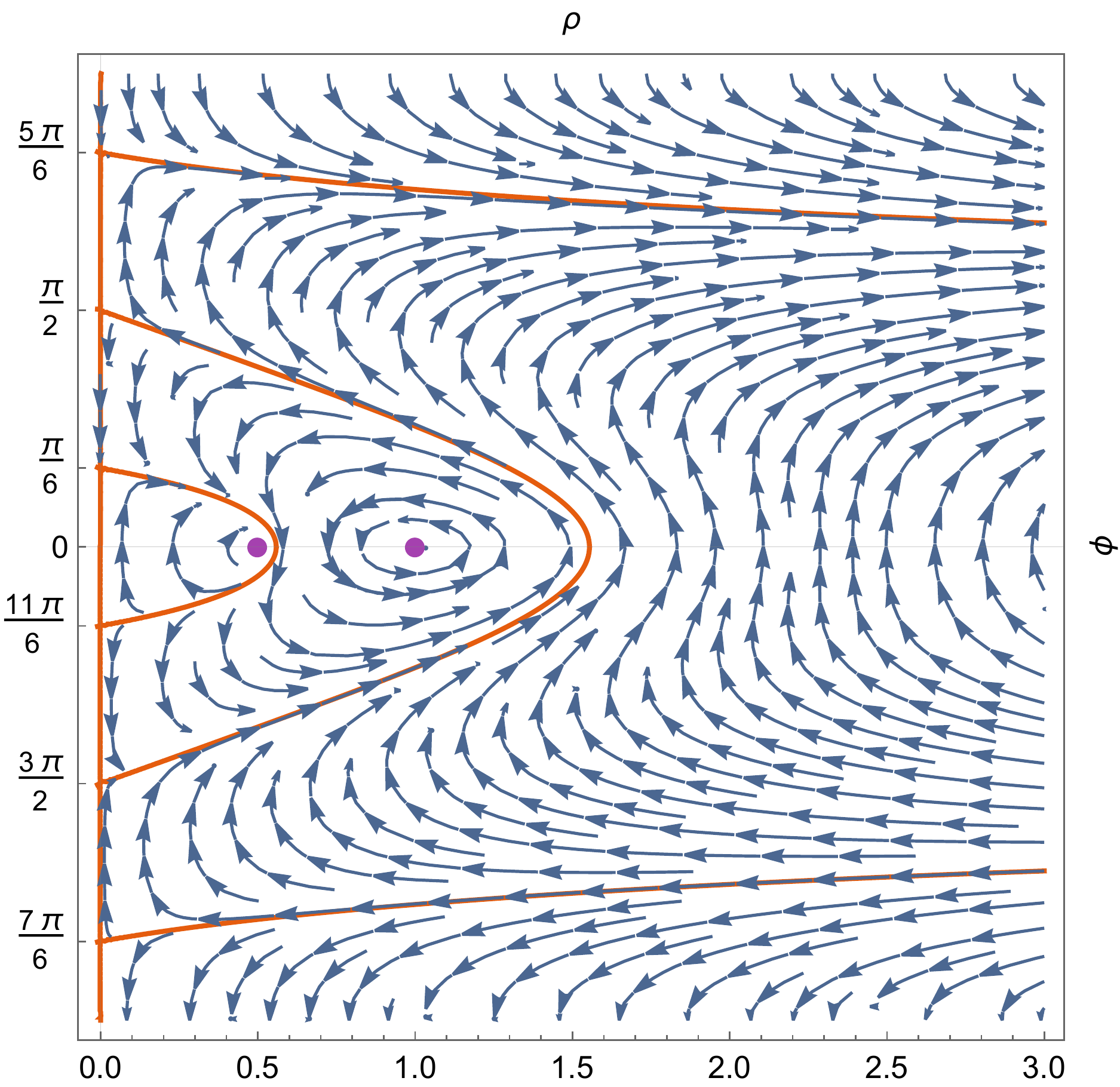}
         \caption{Reissner--Nordstr\"om BH}
         \label{fig:stokes-diag-polar-rn}
     \end{subfigure}
      \captionsetup{width=.9\textwidth}
       \caption{Stokes diagrams in polar coordinates;  
    $\star\dif\Re[z(\rho, \phi)]$--blue vector fields, and 
    the yellow-brown stream corresponds to the exact Stokes lines. 
    The parameter $k=1/2$ is adopted in the RN BH.}
        \label{fig:stokes-diag-polar}
\end{figure}

Alternatively, to fix the angles between any two adjacent Stokes lines emitting from the critical point,
we can also rewrite Eq.\ \eqref{eq:stokes-field} in the polar coordinates $\{\rho,\phi\}$
\begin{equation}
\star\dif \Re[z]= \Im\left[\rho \me^{\mi \phi}/f(\rho \me^{\mi \phi})\right]   \dif \rho
+ \Re\left[ \me^{\mi \phi}/f(\rho \me^{\mi \phi})\right]\dif \phi.
\end{equation}
The angle relationship can then be directly visualized from the diagram in Fig.\ \ref{fig:stokes-diag-polar}, i.e.,
the intersection of the Stokes lines with $\rho = 0$ corresponds to the incident/exit angle of each line.

From an analytical point of view,
the Stokes line of the Schwarzschild BH in polar coordinates becomes
\begin{equation}
1+\rho ^2-2 \rho  \cos (\phi )=\exp \left[-2 \rho  \cos (\phi )\right]
\end{equation}
where $\rho\ge 0$, and $2\pp\ge\phi\ge 0$, 
see the 
plot in Fig.\ \ref{fig:stokes-diag-polar-sch}.
To fix the angles of the emitting lines from origin,
let us consider the corresponding vector field approaching $\rho\sim0$, 
which gives
\begin{equation}
\frac{\dif\rho}{\dif \phi}\sim \rho  \tan (2 \phi )+O\left(\rho ^2\right).
\end{equation}
The stability condition $\dif\rho/\dif \phi\equiv 0$ implies that $\phi= \pp n/2$
with $n\in \mathbb{Z}$.
The angles of the emitting lines from origin can then be calculated as
\begin{equation}
\phi_{\rm st}=\frac{\pp}{2}n+\frac{\pp}{4},
\qquad n\in \mathbb{Z}.
\end{equation}
Similarly, we can compute the Stokes line 
and emitting angle for the RN BH arriving at 
\begin{equation}
1+4 \rho ^2-4 \rho  \cos (\phi )=
\left[\rho ^2-2 \rho  \cos (\phi )+1\right]^4 
\exp \left[4 \rho  \cos (\phi )\right]
\end{equation}
and
$\phi_{\rm st}=\pp n/3+\pp/6$, 
$n\in \mathbb{Z}$, see the  
plot in Fig.\ \ref{fig:stokes-diag-polar-rn}.

\section{Regular black holes with complex poles}
\label{sec:poles}

\subsection{Trajectory along the Stokes lines}

\begin{figure}[!ht]
     \centering
     \begin{subfigure}[b]{0.445\textwidth}
         \centering
         \includegraphics[width=\textwidth]{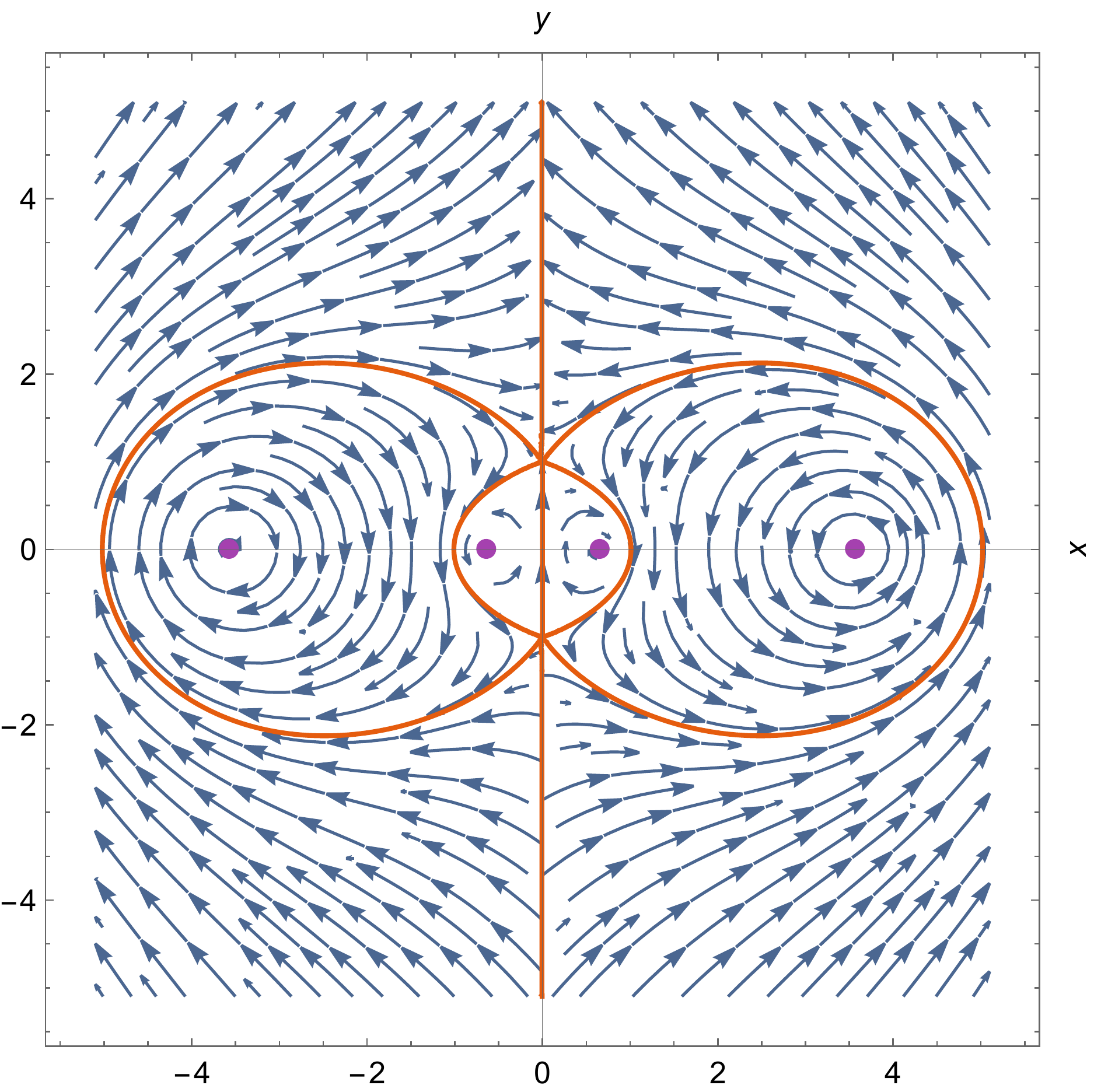}
         \caption{Cartesian coordinates}
     \end{subfigure}
     \begin{subfigure}[b]{0.455\textwidth}
         \centering
         \includegraphics[width=\textwidth]{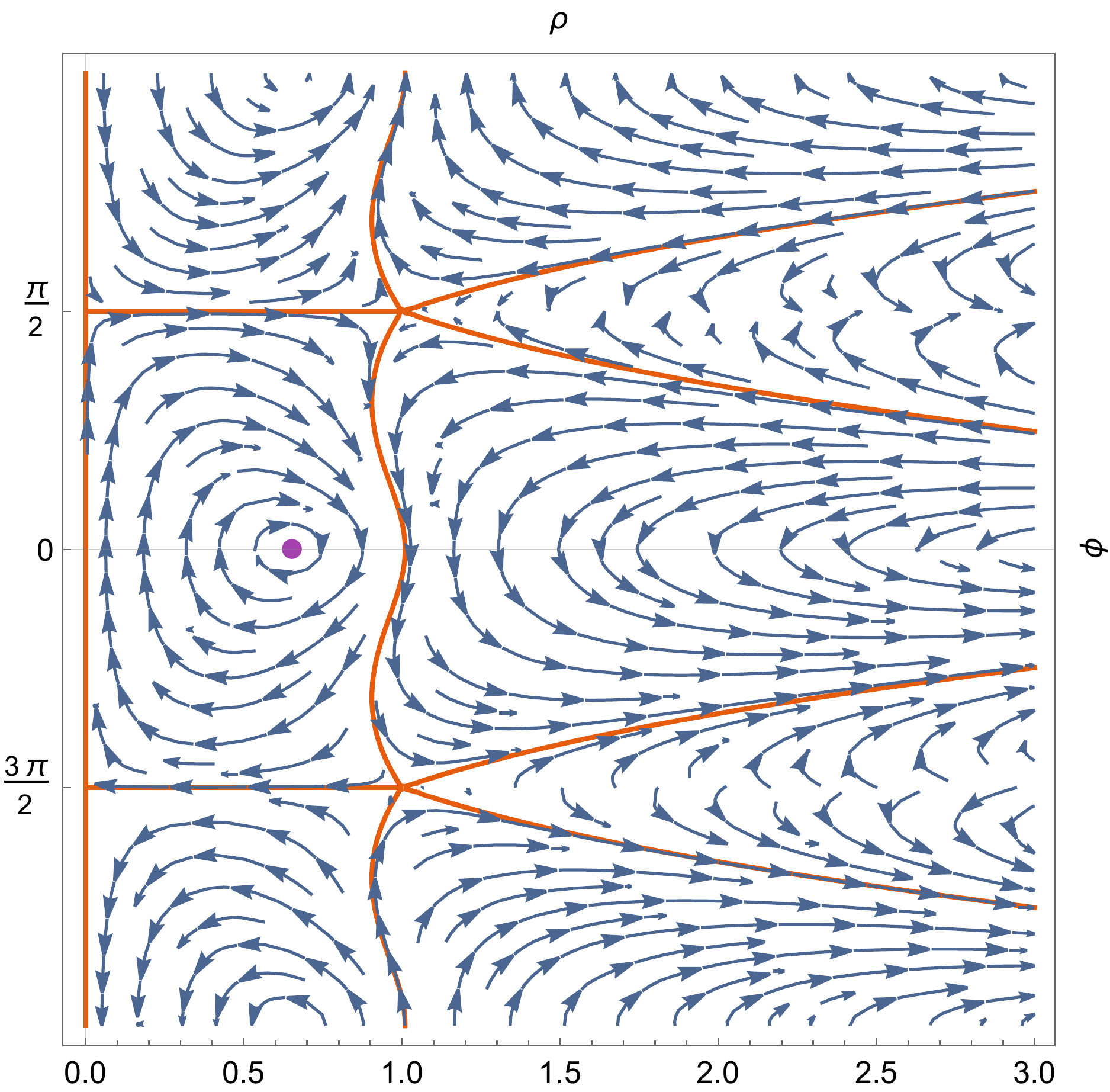}
         \caption{Polar coordinates}
         \label{fig:stokes-diag-polar-bardeen}
     \end{subfigure}
      \captionsetup{width=.9\textwidth}
       \caption{Stokes diagrams for the Bardeen BH; 
    $\star\dif\Re[z(x, y)]$--blue vector fields, and 
    the yellow-brown curves correspond to the Stokes lines,
    where $M=4$. }
        \label{fig:stokes-diag-bardeen}
\end{figure}

Now, let us start with our first example of RBHs, the Bardeen BH \cite{Bardeen:1968non}, which can be generated by a nonlinear magnetic monopole. The Lagrangian of matter generation can be found in Ref.\ \cite{AyonBeato:2000zs,Fan:2016hvf},
\begin{equation}
    \mathcal{L}=\frac{12}{\alpha}\frac{(\alpha\mathcal{F})^{5/4}}{(1+\sqrt{\alpha\mathcal{F}})^{5/2}},\quad \alpha=P^3/M,
\end{equation}
where $\mathcal{F}=F_{\mu\nu}F^{\mu\nu}$ is the field strength, 
and $P$ and $M$ are the magnetic charge and mass parameters, respectively.
We first rescale the shape function by
\begin{equation}
r\to P  r,\qquad M\to \frac{P }{2} M,
\end{equation}
such that it becomes
\begin{equation}
f(r)=1-\frac{M r^2}{\left(r^2+1\right)^{3/2}},
\end{equation}
where all variables and parameters are dimensionless.
The Stokes diagrams are displayed in 
Fig.\ \ref{fig:stokes-diag-bardeen}. 

Note that there are two critical points $\pm \mi$, 
i.e., the singularities of the Bardeen BH are stretched from zero to the pure imaginary axis compared with the RN BH, see Fig.\ \ref{fig:stokes-diag-rn}. 
This can also be understood from the 
Weyl curvature $W$,
which
is regular at zero but divergent at $\pm \mi$, i.e.,
$\pm \mi$ are the poles of the seventh-order,
\begin{equation}
W \sim\pm \frac{75 \mi M^2}{128 P^4 (r\mp \mi)^7}+O[(r\mp i)^{-6}].
\end{equation}
Therefore, $\pm \mi$ are the singular points of the master equation according to the effective potential, Eq.\ \eqref{eq:potential}.
Meanwhile, they are regular singular points because 
\begin{equation}
(r\pm\mi) p=-\frac{3}{2}+O\left[(r\pm \mi)\right] \quad
\text{and} \quad
(r\pm\mi)^2 q=\mp\frac{3}{2} \mi (r\pm \mi)+O\left[(r\pm \mi)^{3/2}\right]
\end{equation}
are regular at the corresponding points, where $p$ and $q$ are the coefficients in Eq.\ \eqref{eq:coefficients} of the master equation for the Bardeen BH.
Furthermore, 
around these two critical points, there is an approximation of the tortoise coordinate
\begin{equation}
z\sim -\frac{4}{5M}\left(1\mp \mi\right) (r\mp \mi)^{5/2},
\end{equation}
and thus, there are five lines emitting from each critical point \cite{Fedoryuk:1993aa}.
The polar diagram in Fig.\ \ref{fig:stokes-diag-polar-bardeen} shows the angles of the Stokes lines passing through the zero point. 
Because $r=0$ is no longer a critical point, the angles at this point are trivial $\pp$, 
the monodromy of the asymptotic solutions around $r=0$ is trivial. 

\begin{figure}[!ht]
     \centering
     \begin{subfigure}[b]{0.45\textwidth}
         \centering
         \includegraphics[width=\textwidth]{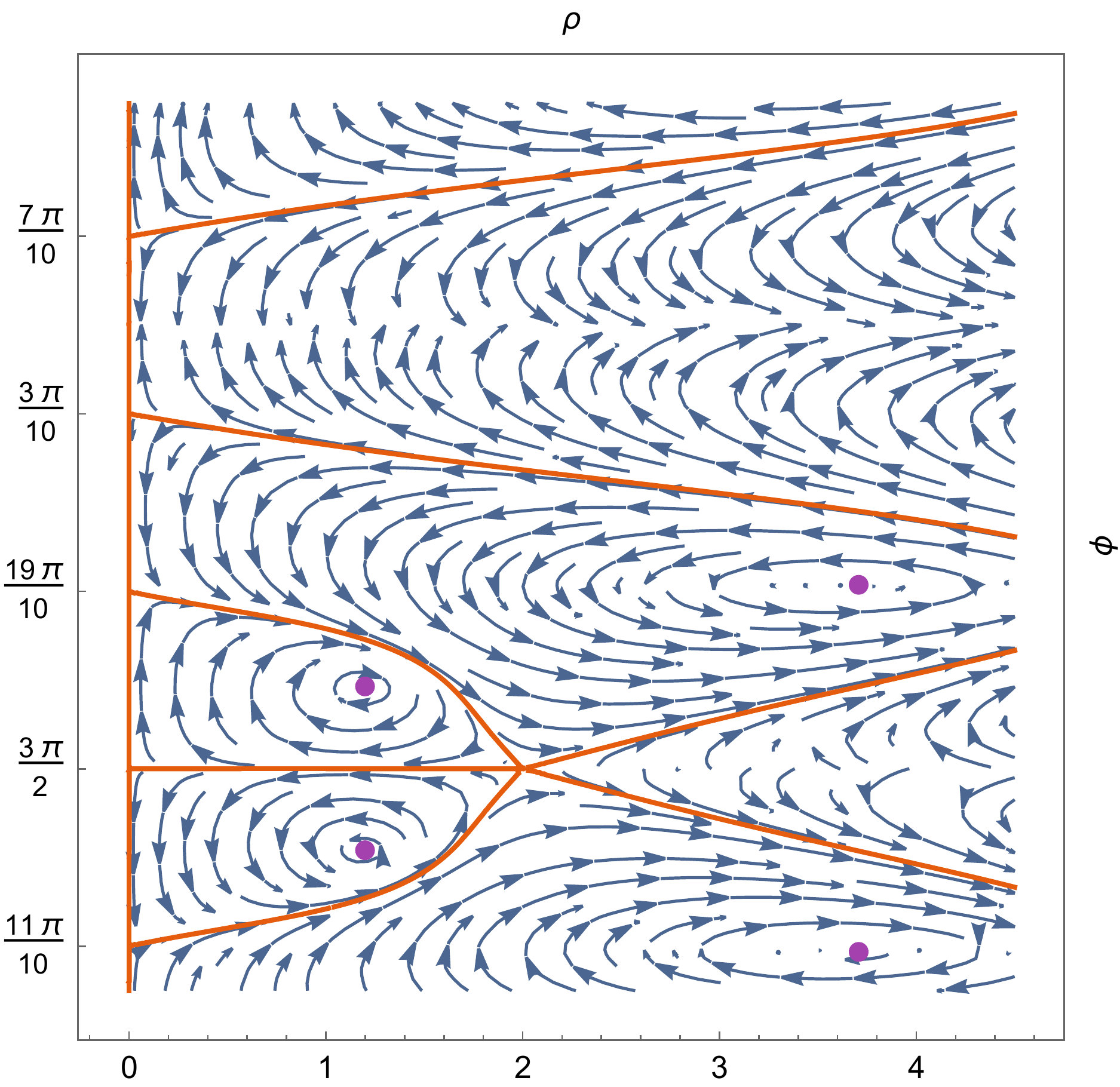}
         \caption{Upper point}
         \label{fig:bardeen-up}
     \end{subfigure}
     \begin{subfigure}[b]{0.45\textwidth}
         \centering
         \includegraphics[width=\textwidth]{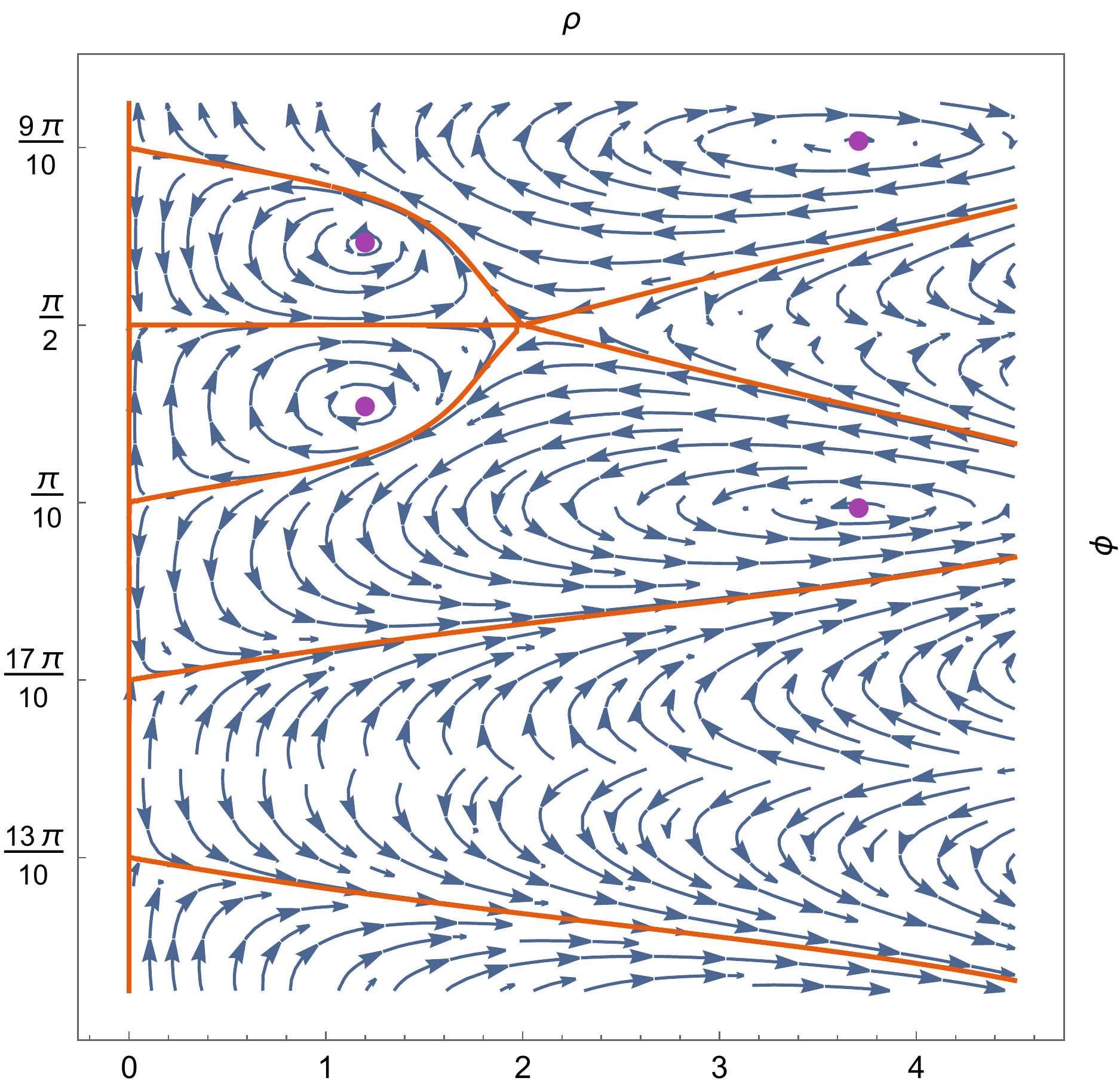}
         \caption{Lower point}
         \label{fig:bardeen-down}
     \end{subfigure}
      \captionsetup{width=.9\textwidth}
        \caption{Stokes diagrams for the Bardeen BH in polar coordinates with starting points at each critical point.
          }
        \label{fig:polar-diag-bardeen-right}
\end{figure}

To obtain information on the angles around the critical points $\pm \mi$,
we should apply the polar coordinates with the starting point at each critical point separately.
Polar diagrams with starting points at $r=\mi$  and $r=-\mi$ are shown in Fig.\ \ref{fig:polar-diag-bardeen-right}.
The angle between adjacent lines can be computed as $2\pp/5$.
Meanwhile, 
it must be emphasized that
the line $\phi=\pp/2$ emits from the lower point,
whereas $\phi=3\pp/2$ emits from the upper point.
This information will help us determine a closed contour, which is applied to calculate the monodromy.

For convenience of later discussions,
let us denote the origin as $\mathcal{O}$, the upper and lower critical points as $\mathcal{A}$ and $\mathcal{B}$ respectively, and a point close to the right inner horizon as $\mathcal{C}$.

Now, let us start with the master equation at zero located on the Stokes line to calculate the AQNMs. Because the tortoise coordinate is approximately
$z\sim r$ at $r=0$, 
we have
\begin{equation}
   \psi ''(z)+  \left[\omega ^2-\frac{l (l+1)}{z^2}\right]\psi (z)=0,
\end{equation}
which gives the solution around $\mathcal{O}$
\begin{equation}
\label{eq:origin}
    \psi_{\mathcal{O}}=Q_+ \sqrt{2\pp \omega z} J_{\frac{\nu}{2}} (\omega z)
    +Q_- \sqrt{2\pp \omega z} J_{-\frac{\nu}{2}} (\omega z),
\end{equation}
where $\nu=2l+1$, and $Q_\pm$ are two arbitrary constants. Then, considering the asymptotics of the Bessel functions  
\begin{equation}
\sqrt{2\pp\omega z} J_{\pm\frac{\nu}{2} }(\omega z)\sim 2 \cos \left(\omega z-\alpha_\pm\right),\qquad 
\omega z>0
\end{equation}
with $\alpha_\pm= \pp \left(1\pm \nu\right)/4$, we find that at $-\mi \infty$, 
\begin{equation}
    \psi_{-\mi\infty}\sim \left(
    Q_+ \me^{\mi \alpha_+}
    +Q_- \me^{\mi \alpha_-}
    \right)\me^{-\mi \omega z}
\end{equation}
and the condition
\begin{equation}
\label{eq:cond1-bardeen}
    Q_+ \me^{-\mi \alpha_+}
    +Q_- \me^{-\mi \alpha_-}=0,
\end{equation}
where the boundary condition $\psi\to \me^{-\mi z\omega}$ as $z\to \infty$ is used.
To pass through $\mathcal{O}$, we must rotate $\pp$ and hence obtain
\begin{equation}\begin{split}
    \psi_{\mathcal{\hat{O}}}&=\me^{2\mi \alpha_+} Q_+ \sqrt{2\pp \omega z} J_{\frac{\nu}{2}} (\omega z)
    +\me^{2\mi \alpha_-}
    Q_- \sqrt{2\pp \omega z} J_{-\frac{\nu}{2}} (\omega z)\\
 &   \sim \left(
 Q_+ \me^{\mi \alpha_+}
 +Q_- \me^{\mi \alpha_-}
 \right) \me^{\mi \omega z}
 +
  \left(
  Q_+ \me^{3\mi \alpha_+}
 +Q_- \me^{3\mi \alpha_-}
 \right) \me^{-\mi \omega z},
\end{split}
\end{equation}
which will be matched with the asymptotics at point $\mathcal{A}$
\begin{equation}\begin{split}
  \psi_{\mathcal{A}}& \sim
 A_+ \sqrt{2\pp \omega (z-\mi)} J_{\frac{\nu}{2}} (\omega (z-\mi))
    +A_- \sqrt{2\pp \omega (z-\mi)} J_{-\frac{\nu}{2}} (\omega (z-\mi))\\
    & \sim \left(
 A_+ \me^{\omega-\mi \alpha_+}
 +A_- \me^{\omega-\mi \alpha_-}
 \right) \me^{\mi \omega z}
 +
  \left(
  A_+ \me^{-\omega+\mi \alpha_+}
 +A_- \me^{-\omega+\mi \alpha_-}
 \right) \me^{-\mi \omega z}.
\end{split}
\end{equation}
Here, we use $\hat{\mathcal{O}}$ to denote the position after rotation around $\mathcal{O}$. This notation also applies to the following discussion.
The matching $ \psi_{\mathcal{A}}\leftrightarrow\psi_{\mathcal{\hat{O}}}$ provides two more conditions:
\begin{equation}
\label{eq:cond2-bardeen}
\begin{split}
   Q_+ \me^{\mi \alpha_+}
 +Q_- \me^{\mi \alpha_-}  & = A_+ \me^{\omega-\mi \alpha_+}
 +A_- \me^{\omega-\mi \alpha_-},\\
   Q_+ \me^{3\mi \alpha_+}
 +Q_- \me^{3\mi \alpha_-} & =
   A_+ \me^{-\omega+\mi \alpha_+}
 +A_- \me^{-\omega+\mi \alpha_-}.
 \end{split}
\end{equation}
To pass through $\mathcal{A}$, we also rotate $\pp$ instead of $2\pp$, which provides
\begin{equation}
    \psi_{\mathcal{\hat{A}}}
    \sim \left(
 A_+ \me^{\omega+\mi \alpha_+}
 +A_- \me^{\omega+\mi \alpha_-}
 \right) \me^{\mi \omega z}
 +
  \left(
  A_+ \me^{-\omega+3\mi \alpha_+}
 +A_- \me^{-\omega+3\mi \alpha_-}
 \right) \me^{-\mi \omega z},
\end{equation}
and the asymptotics as $z$ approaches the inner horizon are 
\begin{equation}\begin{split}
  \psi_{\mathcal{C}}& =
 C_+ \sqrt{2\pp \omega (z-\delta)} J_{\frac{\nu}{2}} (\omega (z-\delta))
    +C_- \sqrt{2\pp \omega (z-\delta)} J_{-\frac{\nu}{2}} (\omega (z-\delta))\\
    & \sim \left(
 C_+ \me^{-\mi\omega\delta+\mi \alpha_+}
 +C_- \me^{-\mi\omega\delta+\mi \alpha_-}
 \right) \me^{\mi \omega z}
 +
  \left(
  C_+ \me^{\mi\omega\delta-\mi \alpha_+}
 +C_- \me^{\mi\omega\delta-\mi \alpha_-}
 \right) \me^{-\mi \omega z},
\end{split}
\end{equation}
where $\delta =\mi /(2 T_{\rm H}^-)$, and
$T_{\rm H}^-$ is the ``temperature'' of the inner horizon. We use 
\begin{equation}
   \sqrt{2\pp \omega (z-\delta)} J_{\pm\frac{\nu}{2}}(\omega (z-\delta)) \sim
   2\cos(\omega (z-\delta) + \alpha_\pm),\qquad
   (z-\delta) \omega \ll -1,
\end{equation}
because $\omega (z-\delta)$ is negative on this branch.
Then, the matching $ \psi_{\mathcal{\hat A}}\leftrightarrow\psi_{\mathcal{C}}$
 gives
\begin{equation}
\label{eq:cond3-bardeen}
\begin{split}
   C_+ \me^{-\mi\omega\delta+\mi \alpha_+}
 +C_- \me^{-\mi\omega\delta+\mi \alpha_-}  & = A_+ \me^{\omega+\mi \alpha_+}
 +A_- \me^{\omega+\mi \alpha_-},\\
    C_+ \me^{\mi\omega\delta-\mi \alpha_+}
 +C_- \me^{\mi\omega\delta-\mi \alpha_-} & =
   A_+ \me^{-\omega+3\mi \alpha_+}
 +A_- \me^{-\omega+3\mi \alpha_-}.
 \end{split}
\end{equation}
Subsequently, we will rotate to the branch containing $\mathcal{B}$. We find
\begin{equation}
     \psi_{\mathcal{\hat C}}
     \sim \left(
 C_+ \me^{-\mi\omega\delta+3\mi \alpha_+}
 +C_- \me^{-\mi\omega\delta+3\mi \alpha_-}
 \right) \me^{\mi \omega z}
 +
  \left(
  C_+ \me^{\mi\omega\delta+\mi \alpha_+}
 +C_- \me^{\mi\omega\delta+\mi \alpha_-}
 \right) \me^{-\mi \omega z},
\end{equation}
which should be matched with the asymptotics at $\mathcal{B}$
\begin{equation}
\begin{split}
  \psi_{\mathcal{B}} &=
 B_+ \sqrt{2\pp \omega (z+\mi)} J_{\frac{\nu}{2}} (\omega (z+\mi))
    +B_- \sqrt{2\pp \omega (z+\mi)} J_{-\frac{\nu}{2}} (\omega (z+\mi))  \\
&  \sim     
\left(
B_1 \me^{-\omega+\mi \alpha_+}
+B_2 \me^{-\omega +\mi \alpha_-}
\right)\me^{\mi \omega z}
+\left(
B_1 \me^{\omega-\mi \alpha_+}
+B_2 \me^{\omega -\mi \alpha_-}
\right)\me^{-\mi \omega z}.
\end{split}
\end{equation}
The matching $ \psi_{\mathcal{\hat C}}\leftrightarrow\psi_{\mathcal{B}}$ leads to another two conditions:
\begin{equation}
\label{eq:cond4-bardeen}
\begin{split}
   B_1 \me^{-\omega+\mi \alpha_+}
+B_2 \me^{-\omega +\mi \alpha_-} & = C_+ \me^{-\mi\omega\delta+3\mi \alpha_+}
 +C_- \me^{-\mi\omega\delta+3\mi \alpha_-}, \\
   B_1 \me^{\omega-\mi \alpha_+}
+B_2 \me^{\omega -\mi \alpha_-} &
=C_+ \me^{\mi\omega\delta+\mi \alpha_+}
 +C_- \me^{\mi\omega\delta+\mi \alpha_-}.
 \end{split}
\end{equation}
We then pass to the origin $\mathcal{O}$ again but in a different branch. To distinguish the solution of the master equation with the original one Eq.\ \eqref{eq:origin},  let us denote it as $\widetilde \psi_{\mathcal{ O}}$, i.e.
\begin{equation}
\begin{split}
   \widetilde \psi_{\mathcal{ O}} & =
\widetilde Q_+ \sqrt{2\pp \omega z} J_{\frac{\nu}{2}} (\omega z)
    +\widetilde Q_- \sqrt{2\pp \omega z} J_{-\frac{\nu}{2}} (\omega z)  \\ 
   & \sim 
   \left(\widetilde{Q}_+ \me^{\mi \alpha_+}
   +\widetilde{Q}_- \me^{\mi \alpha_-}
   \right)\me^{\mi \omega z}
   +
   \left(\widetilde{Q}_+ \me^{-\mi \alpha_+}
   +\widetilde{Q}_- \me^{-\mi \alpha_-}
   \right)\me^{-\mi \omega z},
\end{split}
\end{equation}
which should be matched with the asymptotics at $\mathcal{B}$,
$ \psi_{\mathcal{ B}}\leftrightarrow\psi_{\mathcal{\widetilde O}}$
\begin{equation}
\label{eq:cond5-bardeen}
\begin{split}
   \widetilde{Q}_+ \me^{\mi \alpha_+}
   +\widetilde{Q}_- \me^{\mi \alpha_-}  & = B_1 \me^{-\omega+\mi \alpha_+}
+B_2 \me^{-\omega +\mi \alpha_-}, \\
   \widetilde{Q}_+ \me^{-\mi \alpha_+}
   +\widetilde{Q}_- \me^{-\mi \alpha_-} & =
   B_1 \me^{\omega-\mi \alpha_+}
+B_2 \me^{\omega -\mi \alpha_-}.
 \end{split}
\end{equation}
Finally, after rotating $\pp$ around the origin, we arrive at
\begin{equation}
\begin{split}
   \widetilde \psi_{\mathcal{ \hat O}} & =
\me^{2\mi \alpha_+}\widetilde Q_+ \sqrt{2\pp \omega z} J_{\frac{\nu}{2}} (\omega z)
    +\me^{2\mi \alpha_-}\widetilde Q_- \sqrt{2\pp \omega z} J_{-\frac{\nu}{2}} (\omega z)  \\ 
   & \sim 
   \left(\widetilde{Q}_+ \me^{3\mi \alpha_+}
   +\widetilde{Q}_- \me^{3\mi \alpha_-}
   \right)\me^{\mi \omega z}
   +
   \left(\widetilde{Q}_+ \me^{\mi \alpha_+}
   +\widetilde{Q}_- \me^{\mi \alpha_-}
   \right)\me^{-\mi \omega z}.
\end{split}
\end{equation}
Then, as we close the counter and compare it with the change around the outer horizon, we obtain the condition
\begin{equation}
\label{eq:cond6-bardeen}
   \frac{\widetilde{Q}_+ \me^{\mi \alpha_+}
   +\widetilde{Q}_- \me^{\mi \alpha_-}}{ Q_+ \me^{\mi \alpha_+}
    +Q_- \me^{\mi \alpha_-}}
   =\me^{\omega/T_{\rm H}^+}
\end{equation}
where $T_{\rm H}^+$ is the temperature of the outer horizon.
The condition of asymptotic QNMs is computed by combining Eqs.\eqref{eq:cond1-bardeen}, \eqref{eq:cond2-bardeen},\eqref{eq:cond3-bardeen}, \eqref{eq:cond4-bardeen},
\eqref{eq:cond5-bardeen}, and \eqref{eq:cond6-bardeen}, which gives
\begin{equation}
\label{eq:aqnm-bardeen}
    \exp\left[
    \left(
    \frac{1}{T_{\rm H}^-}-2
    \right) \omega
    \right]
    +\frac{\omega }{T_{\rm H}^+}=0
\end{equation}
or 
\begin{equation}
    \omega =\frac{T_{\rm H}^- }{2 T_{\rm H}^--1}
    W_{n}\left[\frac{T_{\rm H}^+}{T_{\rm H}^-}(1-2 T_{\rm H}^- )\right]
\end{equation}
where $n\in \mathbb{Z}$.
This result is completely different from that obtained in Ref.\ \cite{Flachi:2012nv} for two reasons.
First, the origin $r=0$ is no longer a singular point of the Stokes lines in our calculation, and second, the trajectory of the asymptotic solution along the Stokes lines is different from that of the RN BH.

Before we start a new model, let us comment on the  above process. First, the origin point $r=0$ is a common point of the Stokes lines for the Bardeen BH (i.e., there is only one incoming line and one outgoing line), although it is a regular singular point of the master equation; 
therefore, when we follow the contour around it, the rotation of the solution is trivial, and thus the multipole number $l$ does not contribute to the spectrum of AQNMs. 

Second, to calculate the monodromy of the solution, we start the asymptotics, Eq.\eqref{eq:origin} at the point $r=0$. In fact, we can start with any point  on the Stokes lines, and the result will be the same.
Nevertheless, in practice, if one begins the analysis at $r=+\mi$, the master equation becomes
\begin{equation}
\label{eq:master-bardeen2}
\left(  \frac{\dif{}^2 }{\dif z^2}+\omega^2
-\frac{V_0}{z^{8/5}}\right)\psi( z)=0,
\end{equation}
where $V_0=3 \sqrt[5]{-1} M^{2/5}/5^{8/5}$, and $
z\sim -4\left(1- \mi\right) (r- \mi)^{5/2}/(5M)$. 
This equation is not solvable because the recursion relation obtained using the Frobenius method is not solvable as a difference equation. 
Therefore, the asymptotic analysis of the solution becomes complicated.

To provide an effective analysis, one can use the perturbation method by considering the following equation with a perturbative parameter $\epsilon$
\begin{equation}
  \left(  \frac{\dif{}^2 }{\dif z^2}+\omega^2
-\frac{V_0}{z^{2+\epsilon}}\right)\psi( z)=0,  
\end{equation}
the zero-order $O(\epsilon^0)$ of which can be applied in the calculation of the monodromy.
Owing to this approximation, a deviation from the result Eq.\eqref{eq:aqnm-bardeen} is allowed. 

Third, there is no  linear dependence of $\omega$ on $n$, which is considerably different from the case of the traditional BHs with singularities at the centre, e.g. Schwarzschild, and RN BHs, see Ref.\ \cite{Motl:2003cd}.

\subsection{Disconnected Stokes lines}

The second example refers to the Hayward BH \cite{Hayward:2005gi}. Though the original model is proposed following Markov's idea based on the limiting curvature conjecture \cite{Markov:1982lcc}, it can also be reparameterized and interpreted in the context of nonlinear electrodynamics\cite{Fan:2016hvf}. We apply the latter in our discussion. The matter Lagrangian reads as\cite{Fan:2016hvf}
\begin{equation}
    \mathcal{L}=\frac{12}{\alpha}
    \frac{(\alpha\mathcal{F})^{3/2}}{(1+(\alpha\mathcal{F}^{3/4}))^2},\qquad
    \alpha=P^3/M
\end{equation}
With the help of the coordinate transformation
\begin{equation}
\label{eq:scale-hayward}
r\to P r,\qquad M\to \frac{M P}{2},
\end{equation}
the shape function  becomes
\begin{equation}
f(r)=1-\frac{M r^2}{r^3+1}.
\end{equation}

\begin{figure}[!ht]
     \centering
     \begin{subfigure}[b]{0.45\textwidth}
         \centering
         \includegraphics[width=\textwidth]{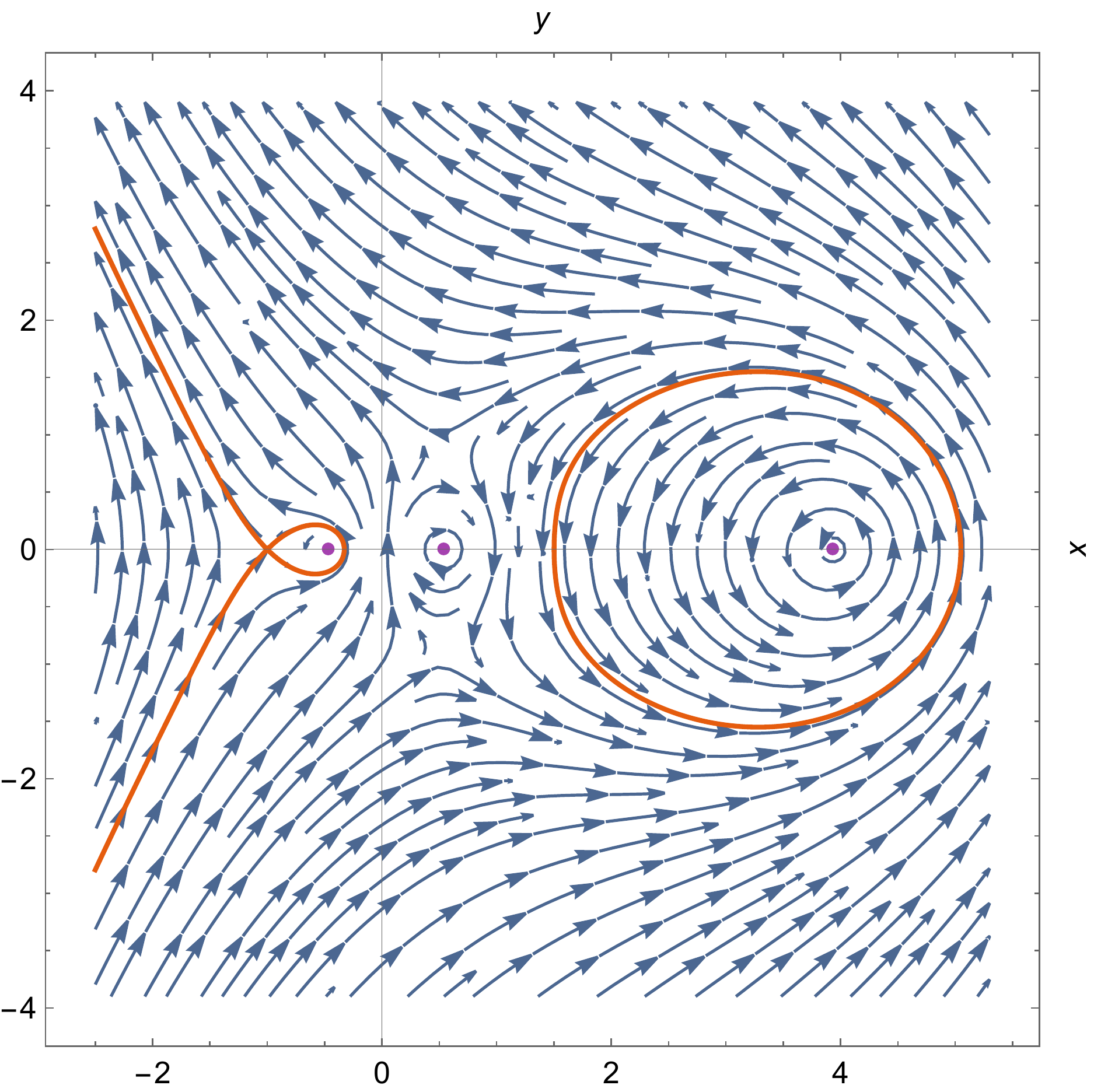}
         \caption{Crossing $-1$.}
         \label{fig:stokes-diag-hayward1}
     \end{subfigure}
     \begin{subfigure}[b]{0.45\textwidth}
         \centering
         \includegraphics[width=\textwidth]{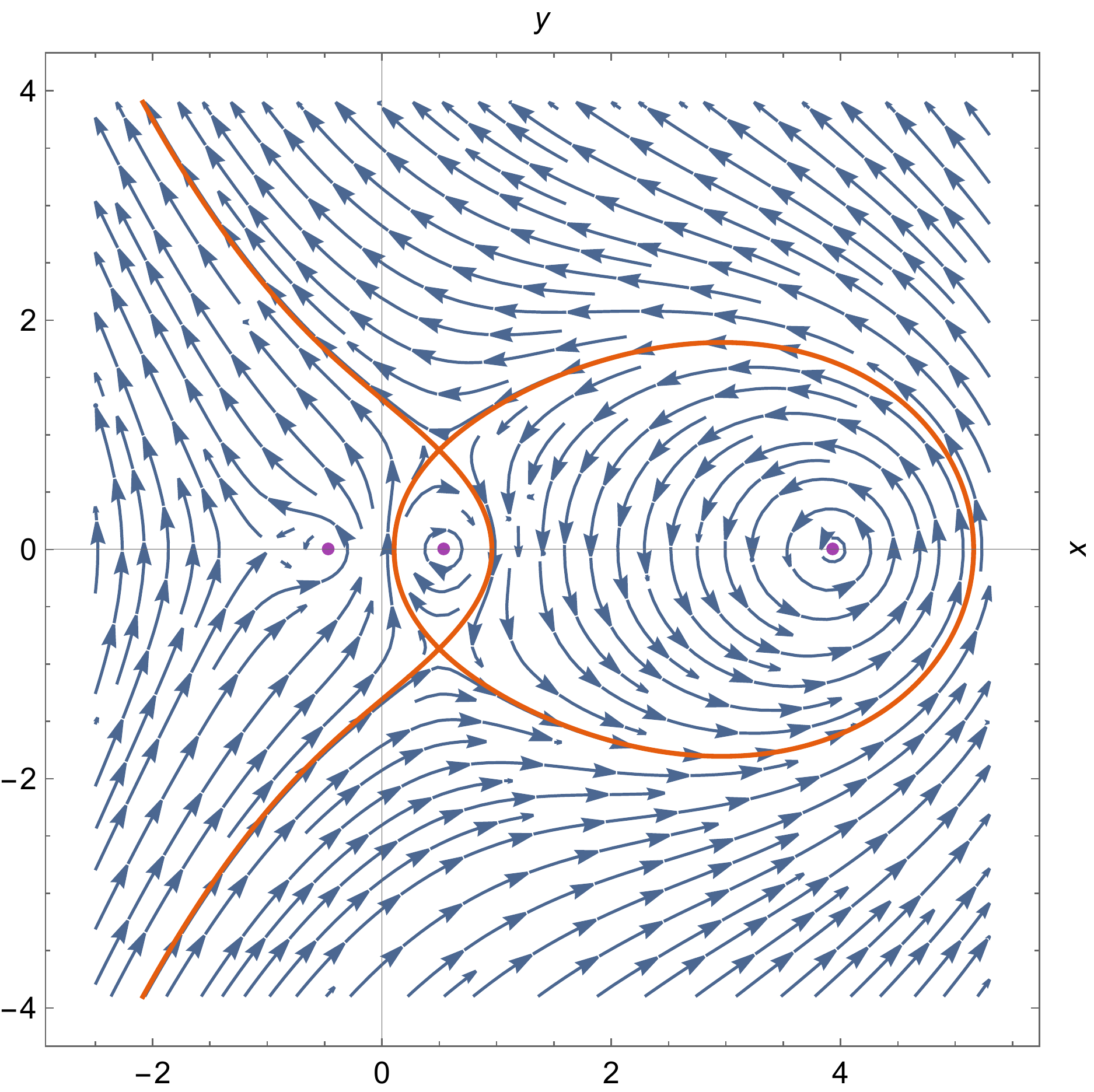}
         \caption{Crossing $(-1)^{1/3}$ and $-(-1)^{2/3}$.}
         \label{fig:stokes-diag-hayward2}
     \end{subfigure}
      \captionsetup{width=.9\textwidth}
        \caption{Stokes diagrams for the Hayward BH; 
    $\star\dif\Re[z(x, y)]$--blue vector fields, and 
    the yellow-brown stream corresponds to the exact Stokes curves,
    where $M=4$. }
        \label{fig:stokes-diag-hayward-true}
\end{figure}

After analytically continuing $r$ into the complex plane, 
we find that the Hayward BH has three horizons. 
The three critical points are $-1$,  $(-1)^{1/3}$, and $-(-1)^{2/3}$,
and the Stokes diagrams are  shown in Fig.\ \ref{fig:stokes-diag-hayward-true},
where the Stokes lines are separated into two groups.
One crosses the critical point $-1$, and the other crosses both $(-1)^{1/3}$ and $-(-1)^{2/3}$ simultaneously;
however, neither crosses the zero point.
Furthermore, because $r=-1$ corresponds to a temperature with a negative BH radius, 
we calculate the monodromy based on Fig.\ \ref{fig:stokes-diag-hayward2}.
In other words, we follow the Stokes line crossing the points $(-1)^{1/3}$ and $-(-1)^{2/3}$.

Information on the angles at each critical point is shown in Fig.\ \ref{fig:polar-diag-bardeen-right}, i.e., the angle between two adjacent lines is $\pp/2$.
\begin{figure}[!ht]
     \centering
     \begin{subfigure}[b]{0.45\textwidth}
         \centering
         \includegraphics[width=\textwidth]{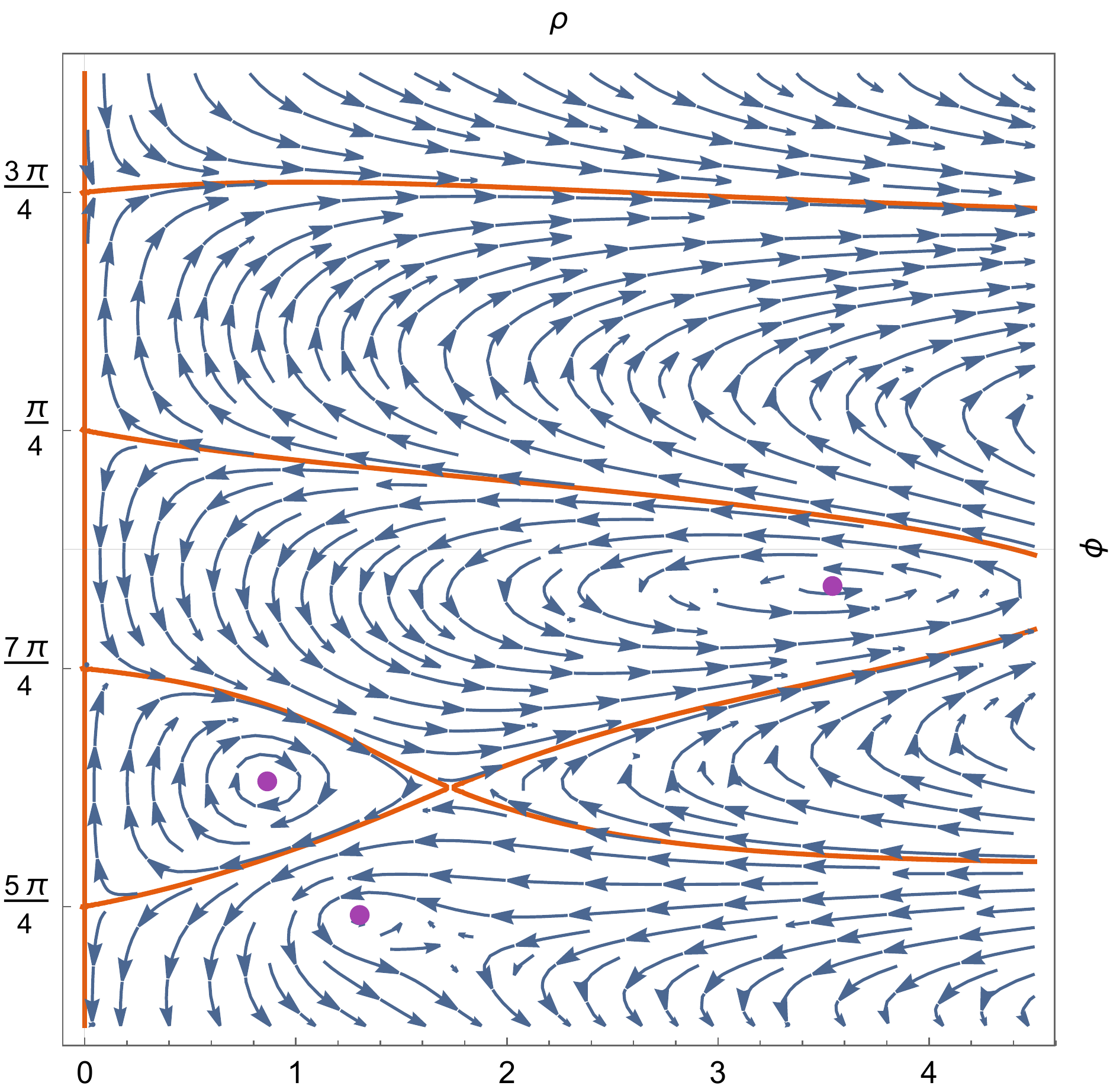}
         \caption{Upper point}
         \label{fig:hayward-up}
     \end{subfigure}
     \begin{subfigure}[b]{0.45\textwidth}
         \centering
         \includegraphics[width=\textwidth]{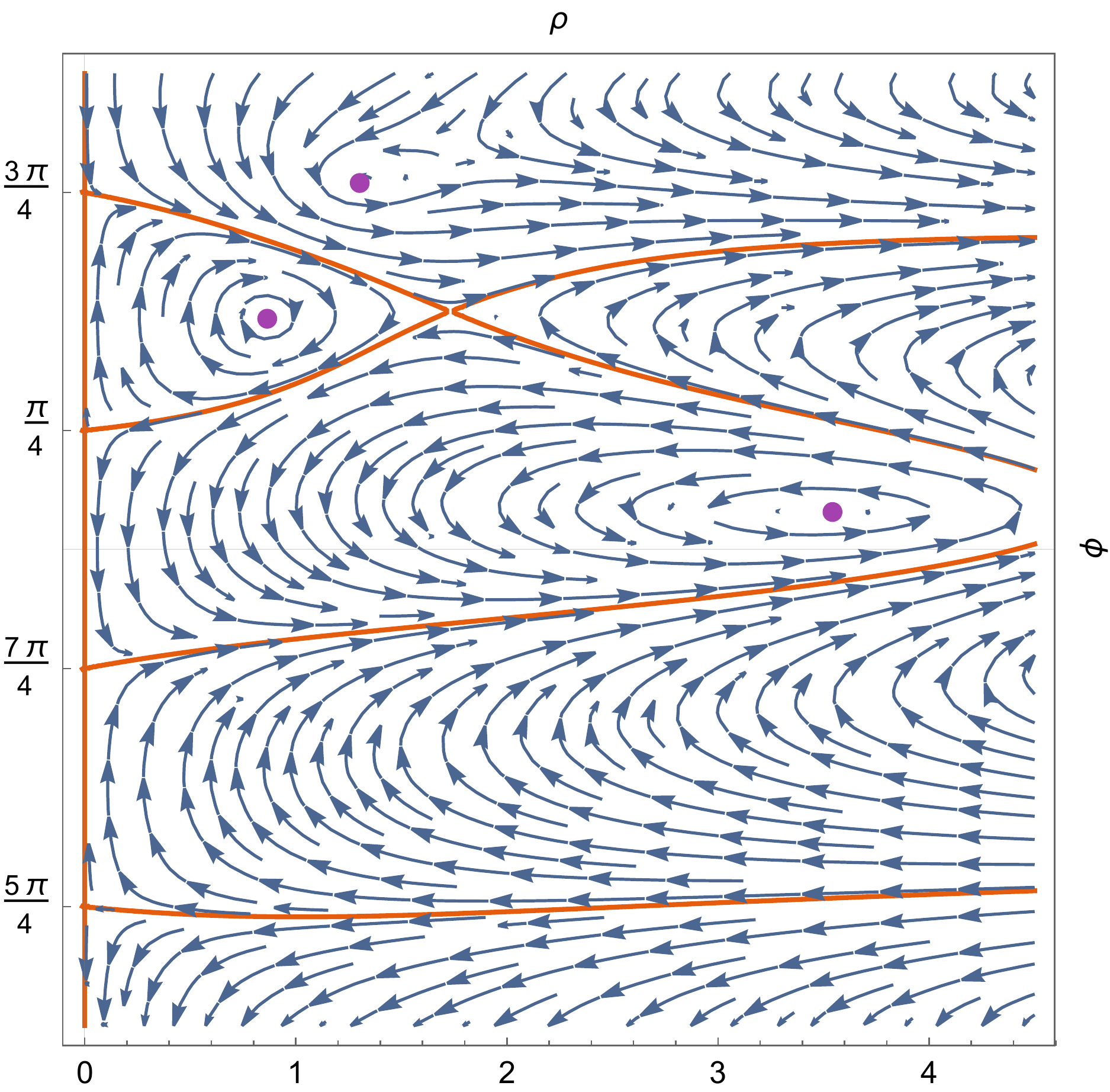}
         \caption{Lower point}
         \label{fig:hayward-down}
     \end{subfigure}
      \captionsetup{width=.9\textwidth}
        \caption{Stokes diagrams of the Hayward BH in polar coordinates, with starting points at each critical point.
          }
        \label{fig:polar-diag-hayward-right}
\end{figure}
The tortoise coordinate around $(-1)^{1/3}$ and $-(-1)^{2/3}$ are, respectively,
\begin{equation}
z \sim -\frac{3 }{2 M} \left[r-(-1)^{1/3}\right]^2,\quad
\text{and}\quad
z \sim -\frac{3 }{2 M} \left[r+(-1)^{2/3}\right]^2.
\end{equation}
Because $r=0$ is not a point on the Stokes line, we start with the lower critical point $r=(-1)^{1/3}$. The master equation then becomes
\begin{equation}
\label{eq:master-hayward}
\left(  \frac{\dif{}^2 }{\dif z^2}+\omega^2
-\frac{V_0}{z^{3/2}}\right)\psi( z)=0,
\end{equation}
where $V_0=-\sqrt[6]{-1} \sqrt{M}/(2 \sqrt{6})$, 
and its coefficient multivalues  owing to $z^{3/2}$.
To perform the asymptotic analysis, we apply the perturbation method based on a Bessel-type equation, as mentioned at the end of previous subsection, i.e. we recast the master equation with
\begin{equation}
\label{eq:perturbation}
\left(  \frac{\dif{}^2 }{\dif z^2}+\omega^2
-\frac{V_0}{z^{2+\epsilon}}\right)\psi(\epsilon, z)=0,
\end{equation}
and expand the wave function as a power series of $\epsilon$, $\psi(\epsilon, z)=\psi_0(z)+\epsilon \psi_1(z)+\dots$, where $\epsilon$ is the perturbative parameter. Then,
we can obtain the perturbative equations order by order
\begin{equation}
\begin{split}
O(\epsilon^0),\quad & \psi_0''(z)+\left(\omega ^2-\frac{V_0}{z^2}\right)\psi_0(z) =0;\\
O(\epsilon^1),\quad & \psi_1''(z)+ \left(\omega ^2-\frac{V_0}{z^2}\right)\psi_1(z)+\frac{V_0  \ln (z)}{z^2}\psi_0(z)=0;\\
\ldots. \quad &
\end{split}
\end{equation}
Next, to calculate the monodromy, 
we only use the zero-order equation and drop the order subscript of the wave function to simplify the notation. In other words, our starting point is the solution of the zero-order  perturbative equation
\begin{equation}
\label{eq:starting-hayward}
    \psi=B_1 \sqrt{2\pp \omega z} J_{\nu/2}(\omega z)+
    B_2 \sqrt{2\pp \omega z} J_{-\nu/2}(\omega z),
\end{equation}
where $\nu=\sqrt{1+4V_0}$ is a complex number.
Moreover, as with the Bardeen BH, we denote the upper and lower critical points as $\mathcal{A}$ and $\mathcal{B}$, respectively, and  a point close to the middle horizon as $\mathcal{C}$.

From the lower critical point, we can obtain an asymptotic behavior of the wave function in Eq.\ \eqref{eq:starting-hayward},
\begin{equation}
    \psi_{\mathcal{B}}\sim
\left(
B_1 \me^{\mi \alpha_+}
+B_2 \me^{\mi \alpha_-}
\right)\me^{-\mi \omega z},
\end{equation}
and the condition
\begin{equation}
    B_1 \me^{-\mi \alpha_+}
+B_2 \me^{-\mi \alpha_-}=0,
\end{equation}
where $\alpha_\pm=\pp(1\pm \nu)/4$.
Then, with the matching $ \psi_{\mathcal{\hat B}}\leftrightarrow\psi_{\mathcal{C}}$, the asymptotics after rotating around $\mathcal{B}$ with that around $\mathcal{C}$ give another two  conditions:
\begin{equation}
    \begin{split}
       B_1 \me^{3\mi \alpha_+}
       +B_2 \me^{3\mi \alpha_-} & =C_1 \me^{-\mi \omega \delta +\mi \alpha_+} 
       +C_2 \me^{-\mi \omega \delta +\mi \alpha_-},\\
       B_1 \me^{5\mi \alpha_+}
       +B_2 \me^{5\mi \alpha_-} & =C_1 \me^{\mi \omega \delta -\mi \alpha_+} 
       +C_2 \me^{\mi \omega \delta -\mi \alpha_-},
    \end{split}
\end{equation}
where $\delta =\mi /(2 T_{\rm H}^-)$,  and $T_{\rm H}^-$ is the ``temperature'' of the inner horizon.

With the matching $\psi_{\mathcal{\hat C}}\leftrightarrow\psi_{\mathcal{A}}$,
the asymptotics of the
wave function at $\mathcal{C}$ after rotating around the upper critical point 
with that at $\mathcal{A}$
provide two more conditions:
\begin{equation}
    \begin{split}
       A_1 \me^{\mi \omega \sqrt{3}+\mi \alpha_+}
       +A_2 \me^{\mi \omega \sqrt{3}+\mi \alpha_-} & =C_1 \me^{-\mi \omega \delta +5\mi \alpha_+} 
       +C_2 \me^{-\mi \omega \delta +5\mi \alpha_-},\\
       A_1 \me^{-\mi \omega \sqrt{3}-\mi \alpha_+}
       +A_2 \me^{-\mi \omega \sqrt{3}-\mi \alpha_-} & =C_1 \me^{\mi \omega \delta +3\mi \alpha_+} 
       +C_2 \me^{\mi \omega \delta +3\mi \alpha_-}.
    \end{split}
\end{equation}
where $\sqrt{3}$ originates from a shift in the variable in Eq.\ \eqref{eq:starting-hayward} from the lower critical point to the upper, i.e.\ $z\to z+\sqrt{3}$,
because the difference between the upper critical point and the lower is $\Im\left[(-1)^{1/3}\right]-\Im\left[-(-1)^{2/3}\right]=\sqrt{3}$.

Finally, after closing the contour
and
comparing it with the change around the outer horizon, we arrive at
\begin{equation}
  \frac{A_1 \me^{\mi \omega \sqrt{3}-\mi \alpha_+}
       +A_2 \me^{\mi \omega \sqrt{3}-\mi \alpha_-}}{B_1 \me^{\mi \alpha_+}
       +B_2 \me^{\mi \alpha_-}}
       =\me^{\frac{\omega}{T_{\rm H}^+}},
\end{equation}
where $T_{\rm H}^+$ denotes the temperature of the outer horizon.
Furthermore, combining all the above conditions, we obtain an analytical expression for the AQMNs,
\begin{equation}
\label{eq:aqnms-hayward}
    \me^{\omega /T^+_{\rm H}}=-2 \me^{-  \omega/T^-_{\rm H} } [\cos (\pp  \nu )+1]-2 \cos (\pp  \nu )-1,
\end{equation}
which has a similar form to the result in Ref.\ \cite{Flachi:2012nv} because the track along the Stokes lines is similar to that of RN BH. 
However, there is an essential difference, in our result, the AQNMs should not depend on the multipole number $l$ because $r=0$ is not a singular point of the Stokes lines, and thus rotation of the asymptotic solution around $r=0$ is trivial, and $l$ does not appear in the formula of AQNMs.

\subsection{Universal form of asymptotic quasinormal modes}

\begin{figure}[!ht]
     \centering
     \begin{subfigure}[b]{0.445\textwidth}
         \centering
         \includegraphics[width=\textwidth]{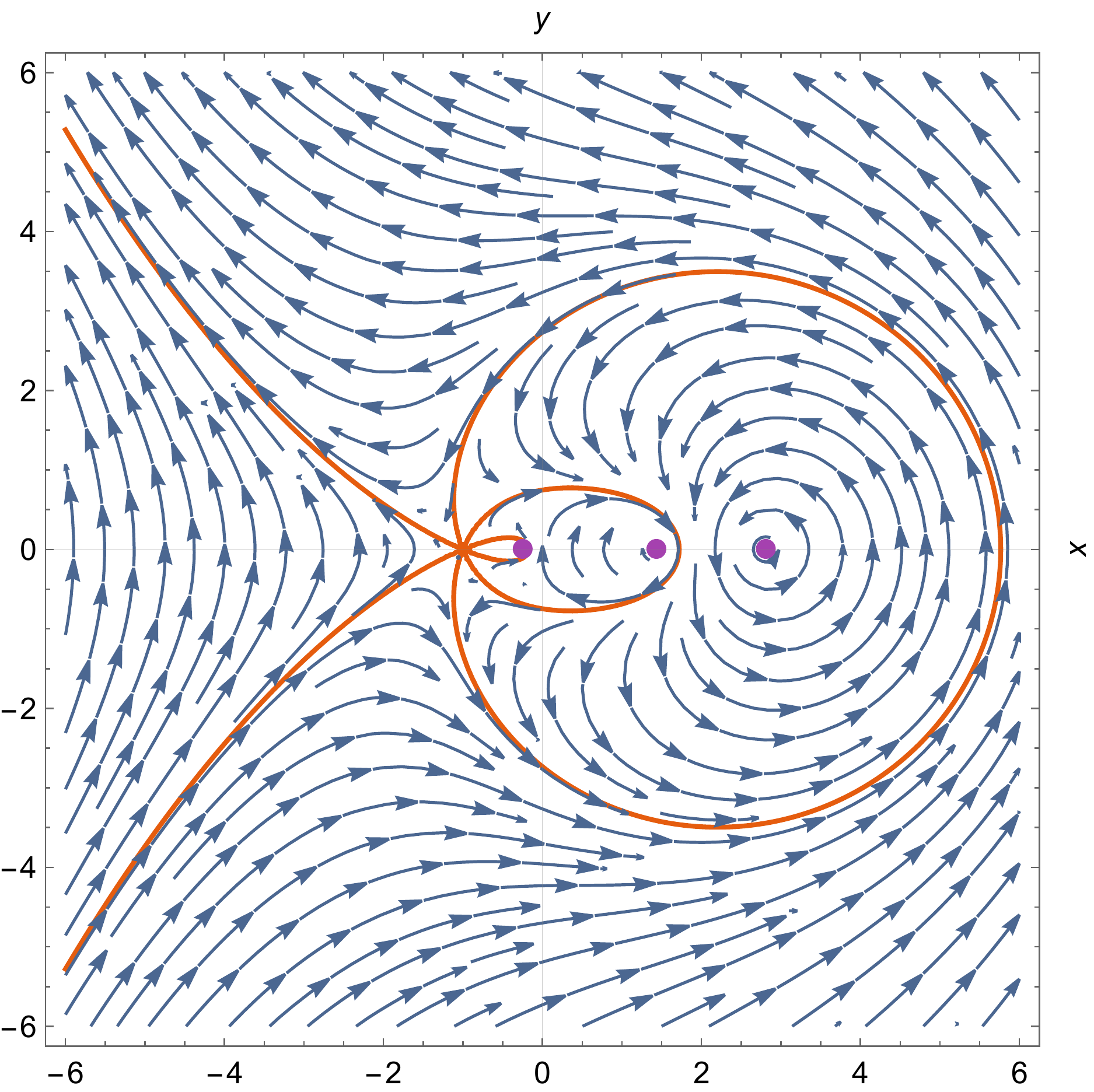}
         \caption{Cartesian coordinates}
         \label{fig:stokes-diag-xy-third}
     \end{subfigure}
     \begin{subfigure}[b]{0.455\textwidth}
         \centering
         \includegraphics[width=\textwidth]{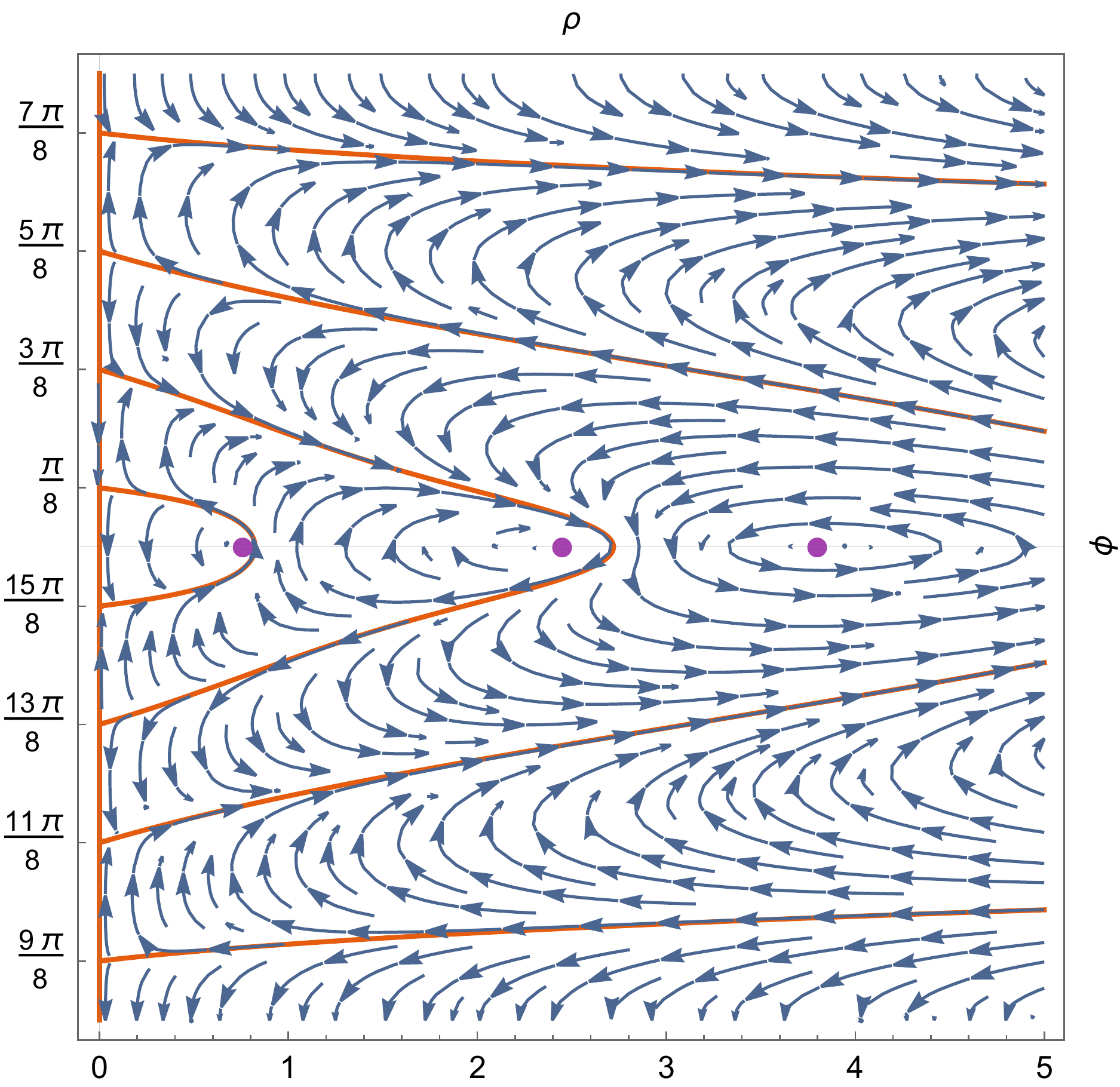}
         \caption{Polar coordinates starting at $r=-1$}
         \label{fig:stokes-diag-polar-third}
     \end{subfigure}
      \captionsetup{width=.9\textwidth}
       \caption{Stokes diagrams for the third class of RBHs; 
    $\star\dif\Re[F(x, y)]$--blue vector fields, and
    the yellow-brown curves correspond to the Stokes lines,
    where $M=7$. }
        \label{fig:stokes-diag-third}
\end{figure}

For the third class of RBHs generalized by the nonlinear electrodynamic source in Ref.\ \cite{Fan:2016hvf},
we simply choose $\mu=3$, yielding the matter Lagrangian \cite{Fan:2016hvf}
\begin{equation}
    \mathcal{L}=\frac{12}{\alpha}
        \frac{\alpha\mathcal{F}}{(1+(\alpha\mathcal{F})^{1/4})^4},\qquad
        \alpha=P^3/M
\end{equation}
and the shape function
\begin{equation}
f=1-\frac{2 M r^{2}}{(P+r)^{3 }},
\end{equation}
and use the transformation $r\to P r$ and $M\to M P/2$
to rescale the variables, such that all the parameters appearing in the shape function are dimensionless.
Furthermore, there is one critical point $r=-1$ and three horizons in this model, see Fig.\ \ref{fig:stokes-diag-third},
and from the critical point, there are eight Stokes lines emitting with adjacent angle of $\pp/4$.

Approaching the critical point, the leading term of the tortoise coordinate is
\begin{equation}
r_*\sim -\frac{(r+1)^4}{4 M},
\end{equation}
and thus the master equation becomes
\begin{equation}
\left(  \frac{\dif{}^2 }{\dif z^2}+\omega^2
-\frac{V_0}{z^{7/4}}\right)\psi( z)=0,
\end{equation}
where $V_0= 3  \sqrt[4]{-M}/(8 \sqrt{2})$.
The same as before, we apply perturbation theory
and take the zero-order solution as the asymptotics of the master equation at the critical point. 
This procedure gives us a similar Bessel-type function. 

On the other side, the closed contour 
starts from the bottom left of Fig.\ \ref{fig:stokes-diag-xy-third}, 
and approaches the critical point
before rotating $\pp/2$ in $r$ (i.e., $2\pp$ in $z$).
Subsequently, it moves toward the middle horizon, 
returns to the critical point, and then finally reaches the top left corner by rotating $\pp/2$ in $r$.

The expression for the AQNMs of this third class is formally the same as that of the Hayward BH, Eq.\ \eqref{eq:aqnms-hayward}, but with a different value of $\nu$. This implies that the matching of $\psi_{\mathcal{\hat C}}\leftrightarrow\psi_{\mathcal{A}}$ in the case of the Hayward BH can be ignored.
Moreover, this reflects the fact that a transformation, such as Eq.\ \eqref{eq:scale-hayward}, should not affect the physical results,
even though the numerical values of the upper and lower critical points depend on the scale transformations.

From the above examples, we clearly see that the AQNMs have the same form as long as the traces of the asymptotic solutions along the Stokes lines are similar. 
Let us analyze one more integrable example.

Inspired by \cite{Peltola:2009jm},
we construct the following shape function:
\begin{equation}
f=\left(1-\frac{P^2}{r^2}\right)^2 \left(\sqrt{1-\frac{P^2}{r^2}}-\frac{2 M}{r}\right),
\end{equation}
where $P$ is interpreted as magnetic charge, as before.
The complete roots of $f(r)=0$ include $\pm P$ and $\pm\sqrt{P^2+4M^2}$,
among which only the positive are physical. The AQNMs of a similar type of RBH have been considered in Refs.\ \cite{Babb:2011ga,Daghigh:2006xg}.

\begin{figure}[!ht]
     \centering
     \begin{subfigure}[b]{0.445\textwidth}
         \centering
         \includegraphics[width=\textwidth]{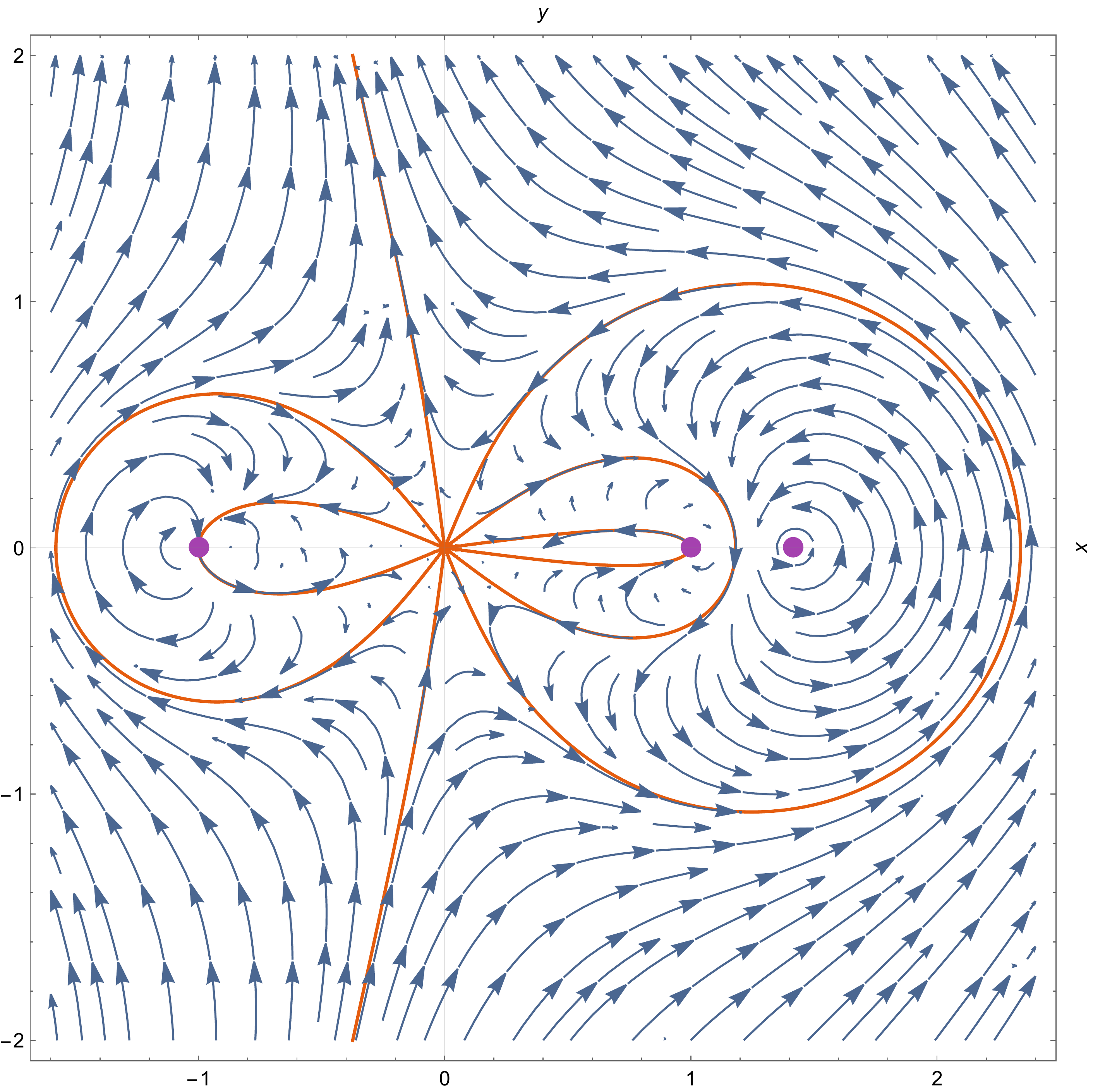}
         \caption{Cartesian coordinates}
         \label{fig:stokesxy-diag-pk}
     \end{subfigure}
     \begin{subfigure}[b]{0.455\textwidth}
         \centering
         \includegraphics[width=\textwidth]{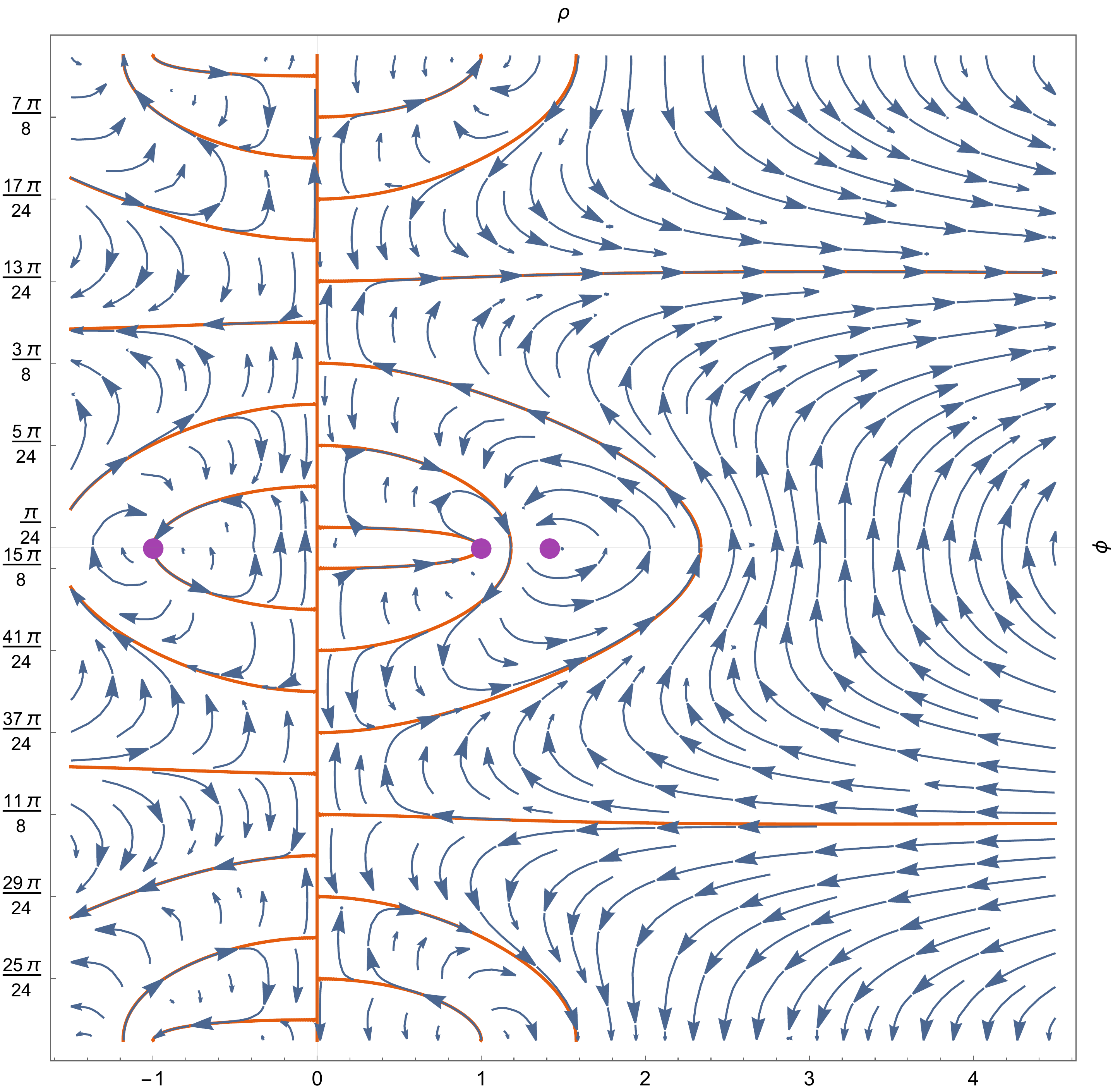}
         \caption{Polar coordinates}
         \label{fig:stokesrhophi-diag-pk}
     \end{subfigure}
      \captionsetup{width=.9\textwidth}
       \caption{Stokes diagrams for the PK BH;
    $\star\dif\Re[F(x, y)]$--blue vector fields, and 
    the yellow-brown curves correspond to the Stokes lines,
    where $M=1/2$ and $P=1$. }
        \label{fig:stokes-diag-pk}
\end{figure}

The physical radius of this BH cannot be less than $P$,
otherwise the metric becomes complex.
Even though the curvature invariants are mathematically divergent at the zero point,
the physical condition $r\ge P$ forbids any test particle from passing into the inner horizon $r_-=P$. 
Therefore, all the curvature invariants are finite in the physical domain $r\in [P,\infty)$.
Moreover, the origin is the only critical point, and the tortoise coordinate can be obtained analytically,
\begin{equation}
2(z+z_0)= \frac{2 r^2}{\sqrt{r^2-1}}+\frac{1}{r^2-1}+\ln\left[\frac{\left(2-r^2\right)^8}{\left(1-r^2\right)^3 \left(r^2+2 \sqrt{r^2-1}\right)^4}\right],
\end{equation}
where $z_0=-1/2+\ln (4)+\mi \pp $, and we use the normalization $r\to 2 M r$ and $P\to 2 M P$ and set $P=1$.
Thus
\begin{equation}
z\sim -\left(1+\mi\right) r^6/12
\end{equation}
which implies that there are twelve Stokes lines emitting from the origin, and
the angle between any two adjacent lines is $\pp/6$, see Fig.\ \ref{fig:stokes-diag-pk}. 
The master equation is 
\begin{equation}
\psi ''(z)+ \left(\omega ^2-\frac{5}{36 z^2}\right)\psi (z)=0.
\end{equation}
Thus, the monodromy relation is similar to that for the RN BH,
\begin{equation}
\me^{\omega/T^+_{\rm H}}=
-\left[
1+2\cos(\pp \nu)
\right]
-2\left[
1+\cos(\pp \nu)
\right]\me^{-\omega/T^-_{\rm H}},
\end{equation}
where $\nu=\sqrt{14}/3$. Substituting the explicit forms of the temperatures of the inner and outer horizons, 
\begin{equation}
T^+_{\rm H}=\frac{2 M^3}{\pp  \left(4 M^2+P^2\right)^2},\qquad
T^-_{\rm H}=0,
\end{equation}
we arrive at
\begin{equation}
\omega/T^+_{\rm H}=
2 \mi \pp  n+
\ln \left[-1-2 \cos \left(\pp\sqrt{14}  /3\right)\right].
\end{equation}

As a step summary, 
we have shown that the analytical form of AQNMs may be universal if the trajectories of the asymptotic solutions along the Stokes lines are similar; however, they should not depend on the multipole number $l$.

\section{Regular black holes  with essential singularities or without singularities}
\label{sec:other}

The models considered in Sec.\ \ref{sec:QNM-scalar}
have a common aspect: the singularities of RBHs still exist (defined as the second types), 
but they are simply moved to the classically forbidden region.
In the current section, we investigate several exotic examples that belong to the first and third types discussed in Sec.\ \ref{sec:singularity}. 

\subsection{Essential singularity as a semi-critical point}
\label{subsec:essential}

Let us express the shape function with an essential singularity at $r=0$ as \cite{Culetu:2014lca,Balart:2014cga}
\begin{equation}
\label{eq:BV-BH-shape}
f(r) = 1- \frac{2M}{r} \sigma(r),\qquad
\sigma(r)= \me^{-\frac{P^2}{2Mr}},
\end{equation}
where $P$ is interpreted as magnetic charge.
The Lagrangian of the matter source as a nonlinear magnetic monopole can be found as
\begin{equation}
\mathcal{L}= \frac{4 P^3 }{\mathcal{F}} \exp\left[-\frac{P^{9/4}}{2^{3/4} M \sqrt[4]{\mathcal{F}}}\right].
\end{equation}
To perform an appropriate analysis, 
we use a dimensionless representation
with the help of the rescaling  transformation \cite{Lan:2020wpv}
\begin{equation}
\label{eq:rescale-BV}
r\to \frac{2 M }{P^2}  r,\qquad
P\to \frac{2 M}{P}.
\end{equation}
The shape function then becomes
\begin{equation}
\label{eq:rescale-BV-shape}
f( r)=1-\frac{P^2 }{ r}\me^{-1/ r}.
\end{equation}
When $r\in \mathbb{R}$, there are two real horizons $r_{\rm H}=-1/W_n\left(-P^{-2}\right)$ with $n=0, -1$ if $P^2> \me$.
Here $W_n(z)$ is the Lambert W function \cite{olver2010nist}.
When $r$ is analytically continued into the complex plane, $n$ is enlarged to all natural numbers $\mathbb{N}$, i.e.,
the roots of $f(r_{\rm H})=0$ are infinitely  many according to Picard's great theorem \cite{shabat2004intro}
because $r=0$ is an {\em essential} singularity of $f(r)$.
Meanwhile, 
all of these roots except the physical horizons are bounded by
$\abs{r_{\rm H}}\le \abs{-1/W_0\left(-P^{-2}\right)}$ and located symmetrically with respect to the real axis because $W^*_n\left(-P^{-2}\right)=W_{-n-1}\left(-P^{-2}\right)$ if $n\neq 0, -1$.

To find the critical points, it is equivalent to calculate the zeros of $r\me^{1/r}$ because 
\begin{equation}
    1/f=\frac{r \me^{1/r}}{r \me^{1/r}-P^2},
\end{equation}
but  $r\me^{1/r}$ does not have any zeros on the complex plane, i.e,
the zero is a Picard exceptional value of $1/f$.
Thus, there are no critical points for this model.
However, the origin $r=0$ is a jump discontinuity,
\begin{equation}
\lim_{r\to 0^+} 1/f=1,\qquad
\lim_{r\to 0^-} 1/f=0,
\end{equation}
which signifies that it may play a special role in the Stokes diagram, 
even though it is not a critical point. The superscript in $0^+$ ($0^-$)
indicates that the path of the limit to $0$ is taken in the right (left) half-plane.
From the perspective of the Weyl curvature 
\begin{equation}
    W=\frac{P^{12} }{48 M^4 }\frac{\me^{-2/r}}{r^{10}}
     [6 (r-1) r+1]^2,
\end{equation}
the jump discontinuity of $r=0$ implies that
\begin{equation}
\lim_{r\to 0^+} W=0,\qquad
\lim_{r\to 0^-} W\to\infty,
\end{equation}
i.e., the Weyl curvature is divergent at the essential singularity $r=0$ as $r$ approaches zero from the left half-plane.
Thus, we dub $r=0$ as the {\em semi-critical} point because it is different from regular points and the critical points considered in Sec.\ \ref{sec:QNM-scalar}.

\begin{figure}[!ht]
         \centering
         \includegraphics[width=.5\textwidth]{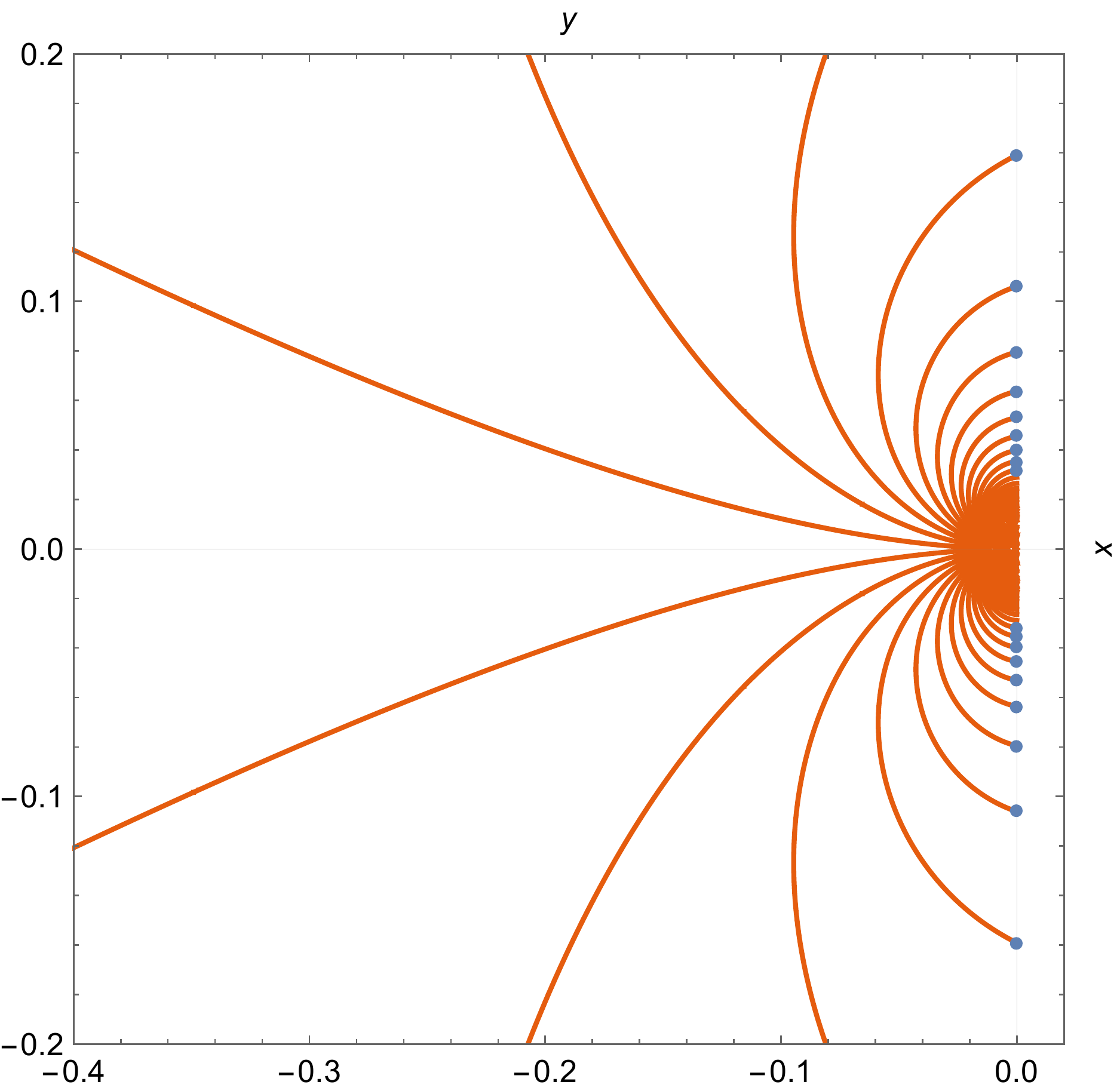}
      \captionsetup{width=.9\textwidth}
       \caption{Stokes diagrams in Cartesian coordinates for the Balart--Vagenas BH close to $0^-$. The yellow-brown curves are Stokes lines, whereas the blue points are their intersections with the longitudinal axis $y$.}
        \label{fig:stokes-diag-bv-zero}
\end{figure}

To perform a quantitative analysis of the Stokes lines, 
we apply the asymptotic relation $1/f(r)\sim - r \me^{1/r}/P^2 $ as $r\to0^-$ and rewrite Eq.\ \eqref{eq:tortoise-coord} as
\begin{align}
z'(r) &\approx- \frac{r}{P^2}  \me^{1/r} \quad\text{with}\quad z(0^-)=0,
\end{align}
which provides the solution
\begin{align}
\label{eq:z-bv-zero}
z(r) & =-\frac{1}{2P^2} \me^{1/r} r (r+1)+\frac{1}{2P^2} {\rm Ei}\left(\frac{1}{r}\right)\sim \frac{\me^{1/r} r^3}{P^2}
\end{align}
where ${\rm Ei}$ denotes the exponential integral \cite{olver2010nist}.
Thus, the Stokes lines $\Re[z]=0$ around $0^-$ are approximately 
\begin{equation}
\label{eq:stokes-line-bv}
y \left(3 x^2-y^2\right) \sin \left(\frac{y}{x^2+y^2}\right)+x \left(x^2-3 y^2\right) \cos \left(\frac{y}{x^2+y^2}\right)=0.
\end{equation}
Along the longitudinal direction, $x=0$, this reduces to
\begin{equation}
0=y^3 \sin \left(\frac{1}{y}\right).
\end{equation}
As a results, we find that there are {\em infinite} Stokes lines quantified by $1/y=n\pp $
with $n\in\mathbb{Z}/\{0\}$
owing to the periodicity of the sine function, see Fig.\ \ref{fig:stokes-diag-bv-zero}.
This phenomenon of infinitely many Stokes lines around the essential singularity $r=0$ is also a reflection of Picard's great theorem.

\begin{figure}[!ht]
         \centering
         \includegraphics[width=.5\textwidth]{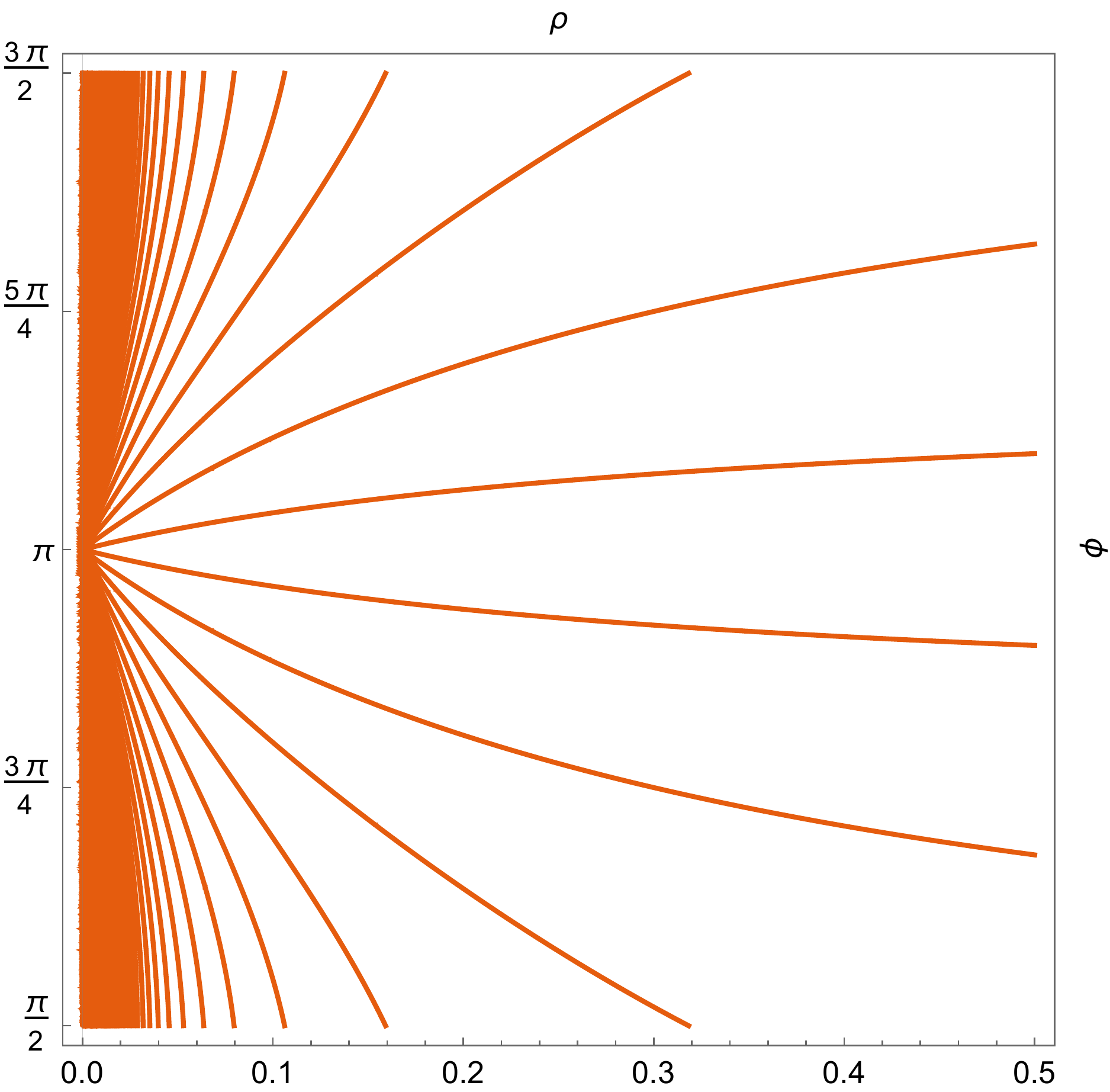}
      \captionsetup{width=.9\textwidth}
       \caption{Stokes diagrams in polar coordinates for the Balart--Vagenas BH close to $0^-$.}
        \label{fig:stokesrhophi-diag-bv-zero}
\end{figure}

To fix the angle between two adjacent Stokes lines at $0^-$, we rewrite Eq.\ \eqref{eq:stokes-line-bv} in the polar system $\{\rho, \phi\}$, which gives
\begin{equation}
    \rho  \cos \left[3 \phi -\frac{\sin (\phi )}{\rho }\right]=0.
\end{equation}
When $\rho$ becomes small, we find that
\begin{equation}
    \sin (\phi )\approx\frac{2 n+1}{2} \pp  \rho\to 0.
\end{equation}
That is, $\phi=\pp$ as $r\to 0^-$. 
In other words, the angles of all Stokes lines emitting from $0^-$ are zero. see Fig.\ \ref{fig:stokesrhophi-diag-bv-zero}.

\begin{figure}[!ht]
     \centering
     \begin{subfigure}[b]{0.445\textwidth}
         \centering
         \includegraphics[width=\textwidth]{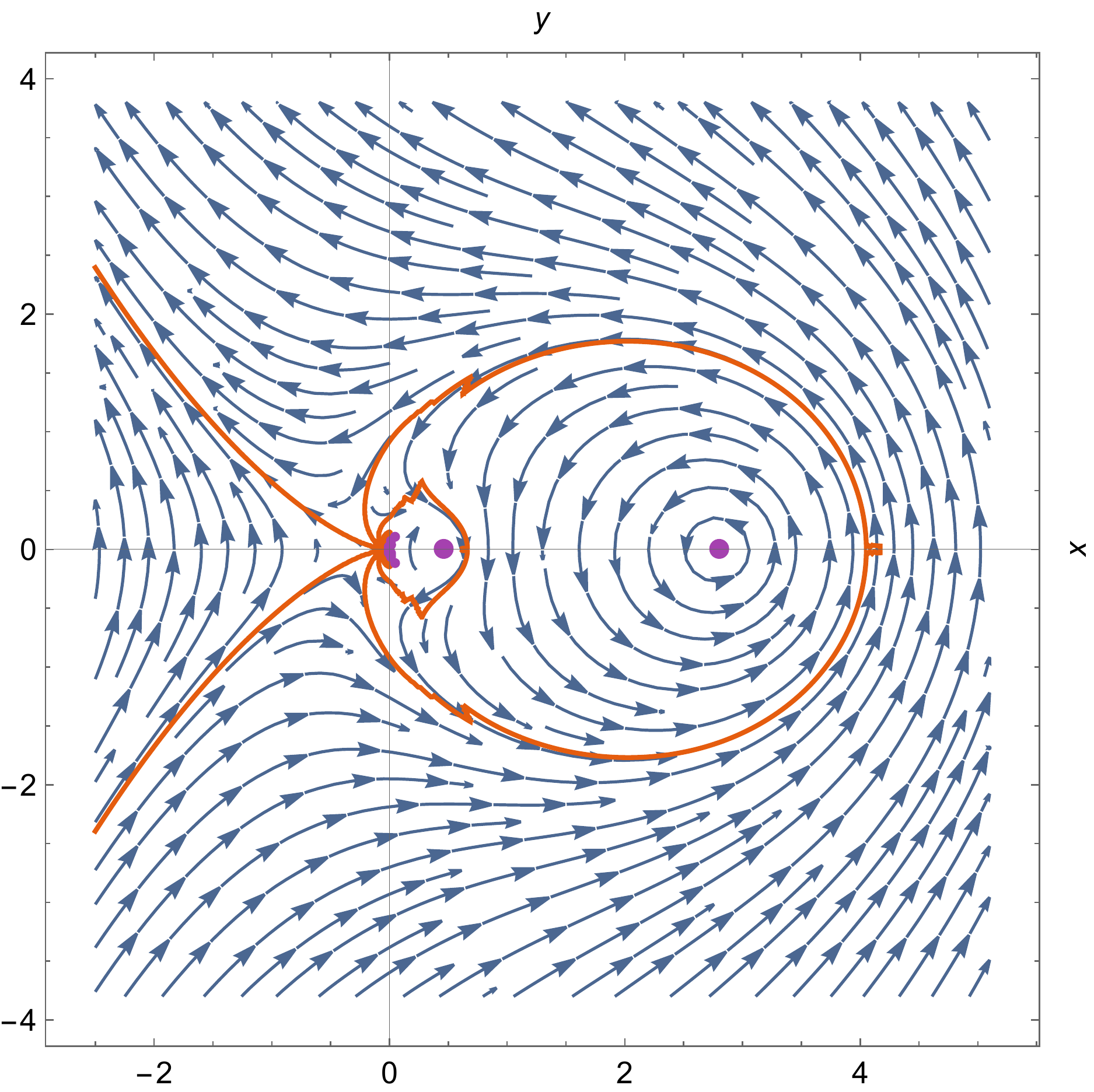}
         \caption{Cartesian coordinates}
         \label{fig:stokesxy-diag-bv}
     \end{subfigure}
     \begin{subfigure}[b]{0.455\textwidth}
         \centering
         \includegraphics[width=\textwidth]{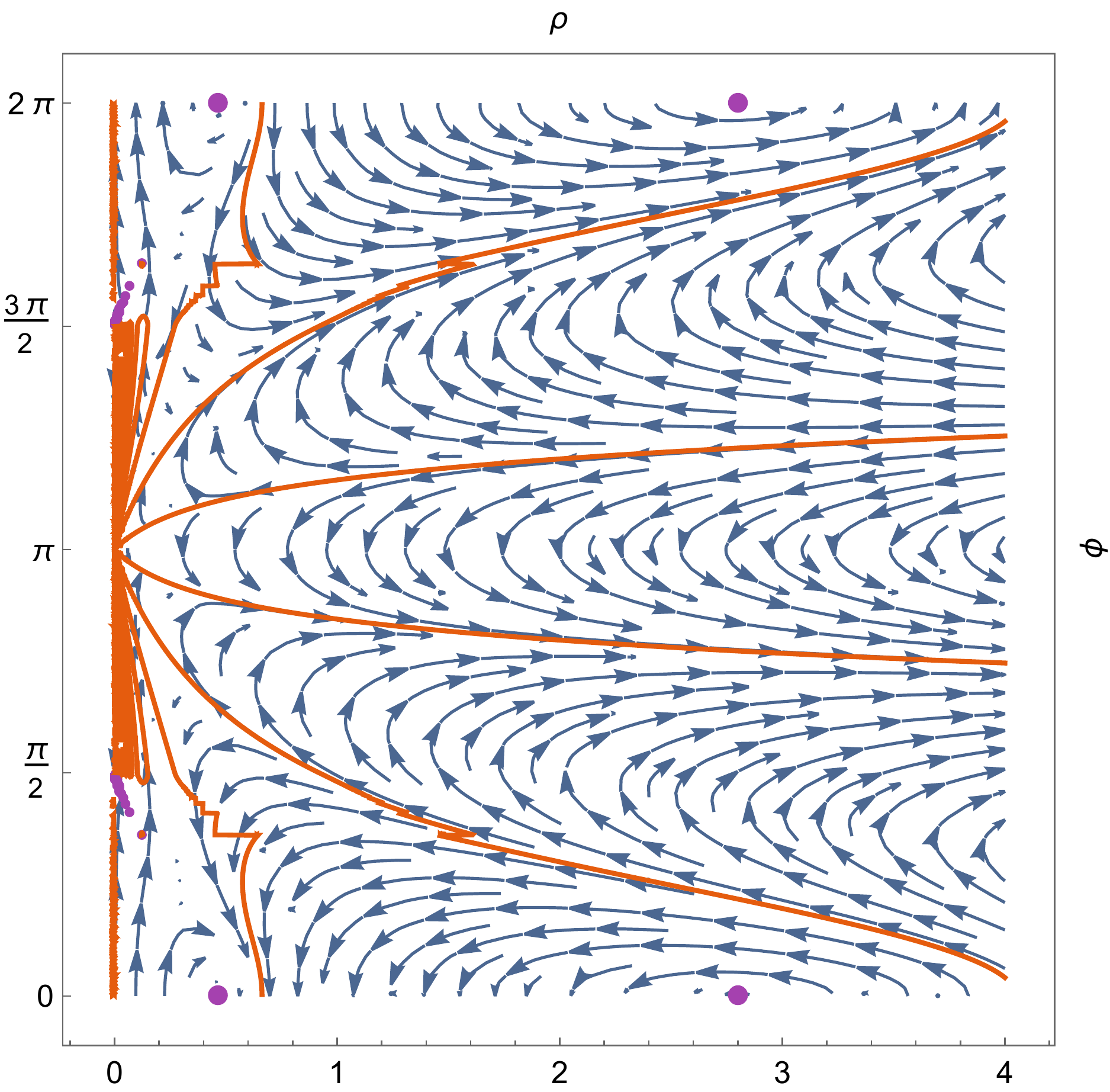}
         \caption{Polar coordinates}
         \label{fig:stokesrhophi-diag-bv}
     \end{subfigure}
      \captionsetup{width=.9\textwidth}
       \caption{Stokes diagrams for the Balart--Vagenas BH;
    $\star\dif\Re[z(x, y)]$--blue vector fields, and
    the yellow-brown curves correspond to the Stokes lines,
    where $P=2$. }
        \label{fig:stokes-diag-bv}
\end{figure}

\begin{figure}[!ht]
     \centering
         \includegraphics[width=0.5\textwidth]{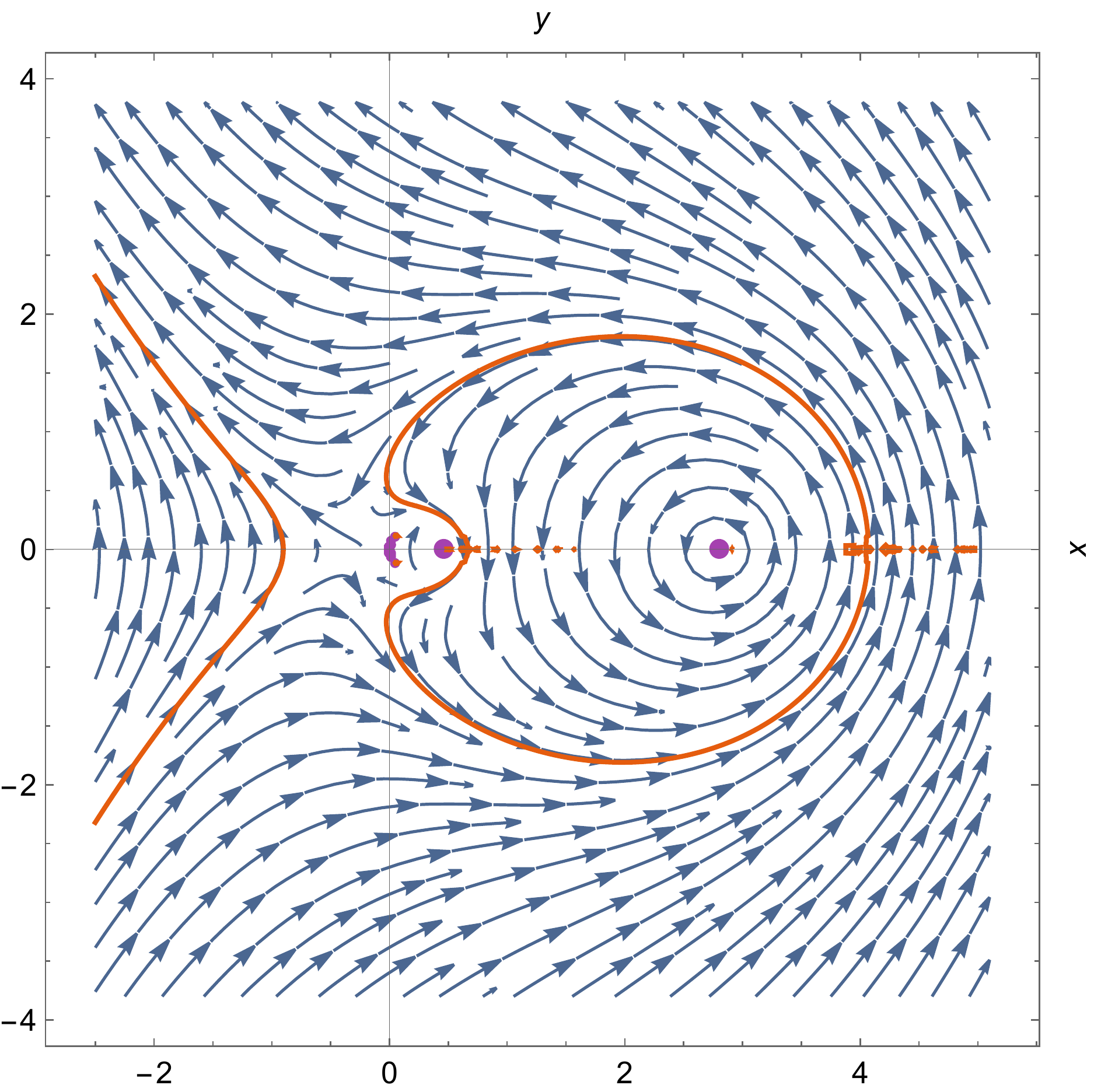}
      \captionsetup{width=.9\textwidth}
       \caption{Improved Stokes lines for the Balart--Vagenas BH
    based on the numeric integral in Eq.\ \eqref{eq:improv-stokes-lines} with $P=2$. }
        \label{fig:stokes-diag-bv-imporv}
\end{figure}

The full Stokes lines are computed using a numeric integral and are depicted in Fig.\ref{fig:stokes-diag-bv}.
Because of the extreme characteristic of the integrand near the singularities, the lines are not smooth at all, as shown in the plot.
Alternatively, we develop a formula based on the Cauchy–Riemann equations \cite{shabat2004intro} to estimate the global feature of Stokes lines in the range far from the origin,
\begin{equation}
\label{eq:improv-stokes-lines}
    0=\int^{x}_0\dif \widetilde x \Re\left[\frac{1}{f(\widetilde x+\mi y)}\right]-
    \int^{y}_0 \dif \widetilde y \lim_{x\to 0}
    \Im\left[\frac{1}{f(x+\mi \widetilde y)}\right],
\end{equation}
which loses its validity as $x\to 0^-$ because an infinite number of complex horizons are located there, see Fig.\ \ref{fig:stokes-diag-bv-imporv}. 

Now we turn to the master equation. First, we obtain the asymptotic form of the potential in Eq.\ \eqref{eq:potential} as $r\to 0^-$
\begin{equation}
    V\sim \frac{P^4 \me^{-2/r}}{r^5},
\end{equation}
and then with the help of the asymptotic relation in Eq.\ \eqref{eq:z-bv-zero},
we can rewrite the master equation as
\begin{equation}
\left[  \frac{\dif{}^2 }{\dif z^2}+\omega^2
+\frac{1}{3 z^2 W_0\left(\frac{1}{3 P^{2/3} \sqrt[3]{-z}}\right)}
\right]\psi(z)=0,
\end{equation}
However, because the angles of the Stokes lines emitting from the origin are zero, 
a trivial monodromy relation is obtained, i.e.\
$1=\exp(\omega/T^+_{\rm H})$, which gives a pure imaginary spectrum of the AQNMs,
$\omega=2\pp\mi n T^+_{\rm H}$.

\subsection{No critical points at all}
\label{sec:NC-Sch}

Next we turn to a RBH model inspired by noncommutative geometry \cite{Nicolini:2005vd},
\begin{equation}
\label{eq:NC-BH-shape}
f(r)= 1-\frac{2M}{r}\sigma,
\qquad
\sigma(r)=\frac{2}{\sqrt{\pp}}\gamma\left(\frac{3}{2},\frac{r^2}{4\epsilon}\right),
\end{equation}
where $\epsilon$ denotes the noncommutative parameter, 
and $\gamma\left(\frac{3}{2},\frac{r^2}{4\epsilon}\right)$ is the lower incomplete gamma function \cite{olver2010nist}.
The original model was proposed by considering the existence of a minimum distance scale.
However, we use the monopole interpretation instead of the original model,
i.e., we replace the parameter $\epsilon$ with $P^2$. 
Thus, the Lagrangian can be obtained as
\begin{equation}
\mathcal{L}=\frac{2 M }{P^3\sqrt{\pp } }\me^{-\frac{\sqrt{\mathcal{F}}}{4 \sqrt{2} P^{5/2}}}.
\end{equation}
Then, with the help of the following transformation \cite{Lan:2020wpv};
\begin{equation}
r\to\frac{2 P  M }{\sqrt{\pp }} r,\qquad
P\to \frac{ P M}{\sqrt{\pp} },
\end{equation}
the shape function becomes
\begin{equation}
\label{eq:shape-nc}
f=1-\frac{2}{P  r} \gamma\left(\frac{3}{2},  r^2\right),
\end{equation}
where the parameter $P$ is bounded by $0.93\gtrsim P>0$,
otherwise there will be not physical horizons.

\begin{figure}[!ht]
     \centering
         \includegraphics[width=\textwidth]{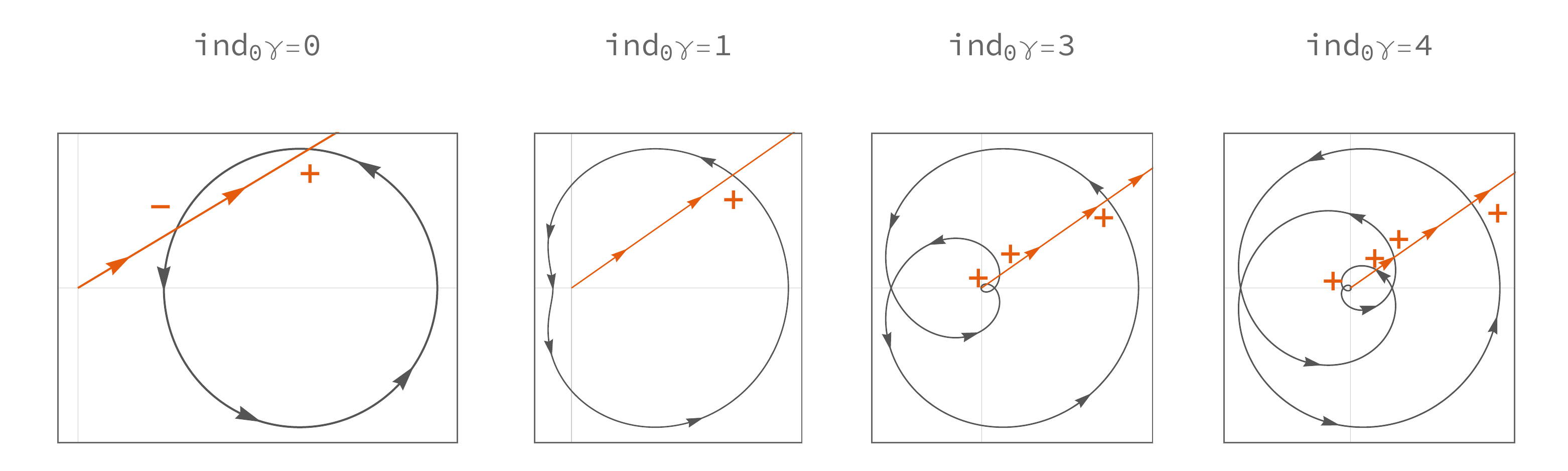}
      \captionsetup{width=.9\textwidth}
       \caption{Winding indices. The images (black closed curves) of circles with radii $\abs{r}=1/2, 1, 3, 4$ from left to right under the map of Eq.\ \eqref{eq:shape-nc}, where $P=1/2$. The yellow-brown rays from zero to infinity are auxiliary lines for calculating the winding indices. The signs ``$\pm$'' correspond to the counter-clockwise and clockwise intersections of the curve with the ray, respectively. }
        \label{fig:winding}
\end{figure}

\begin{figure}[!ht]
     \centering
     \begin{subfigure}[b]{0.445\textwidth}
         \centering
         \includegraphics[width=\textwidth]{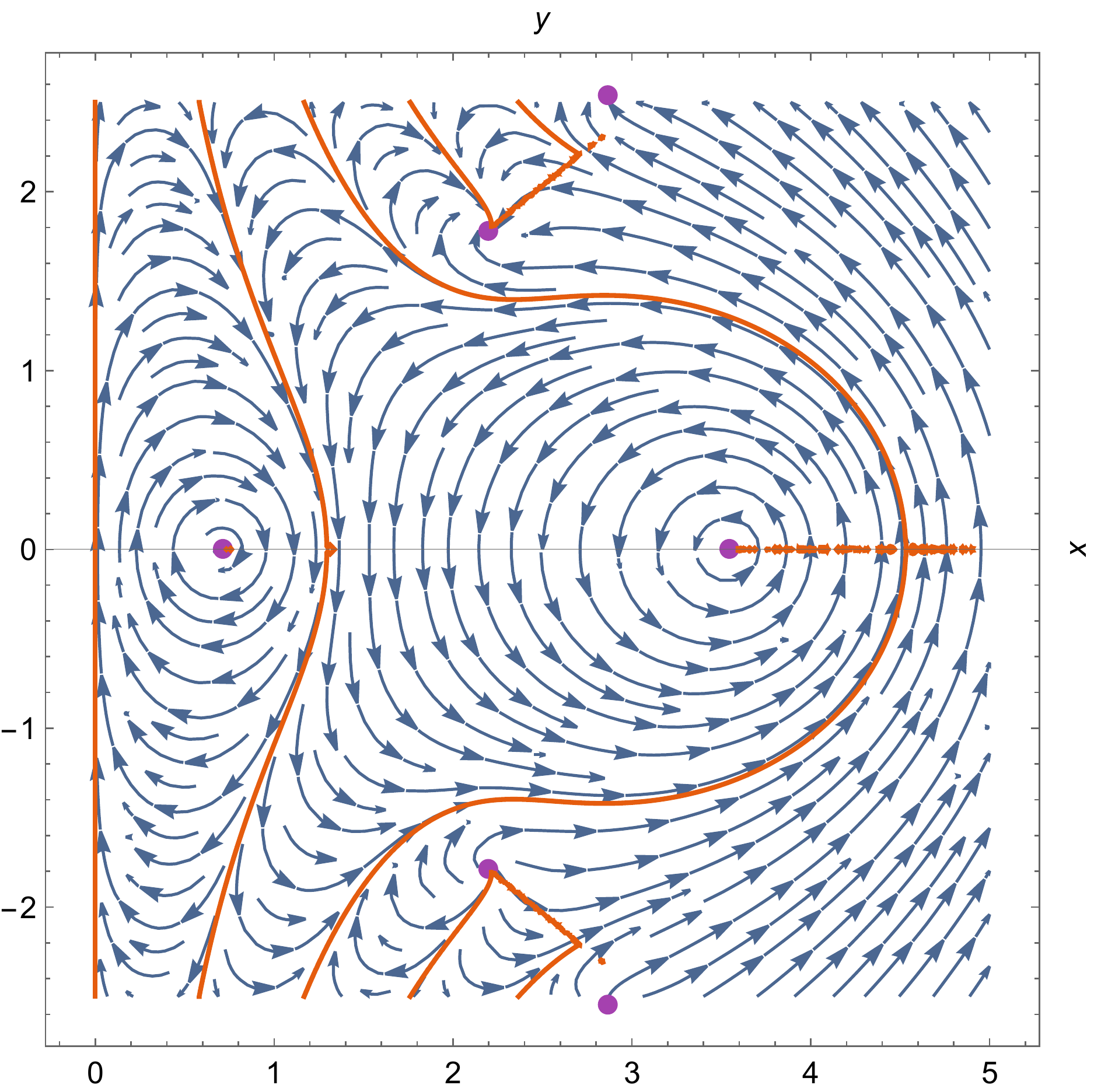}
         \caption{Cartesian coordinates}
         \label{fig:stokes-diag-nc}
     \end{subfigure}
     \begin{subfigure}[b]{0.455\textwidth}
         \centering
         \includegraphics[width=\textwidth]{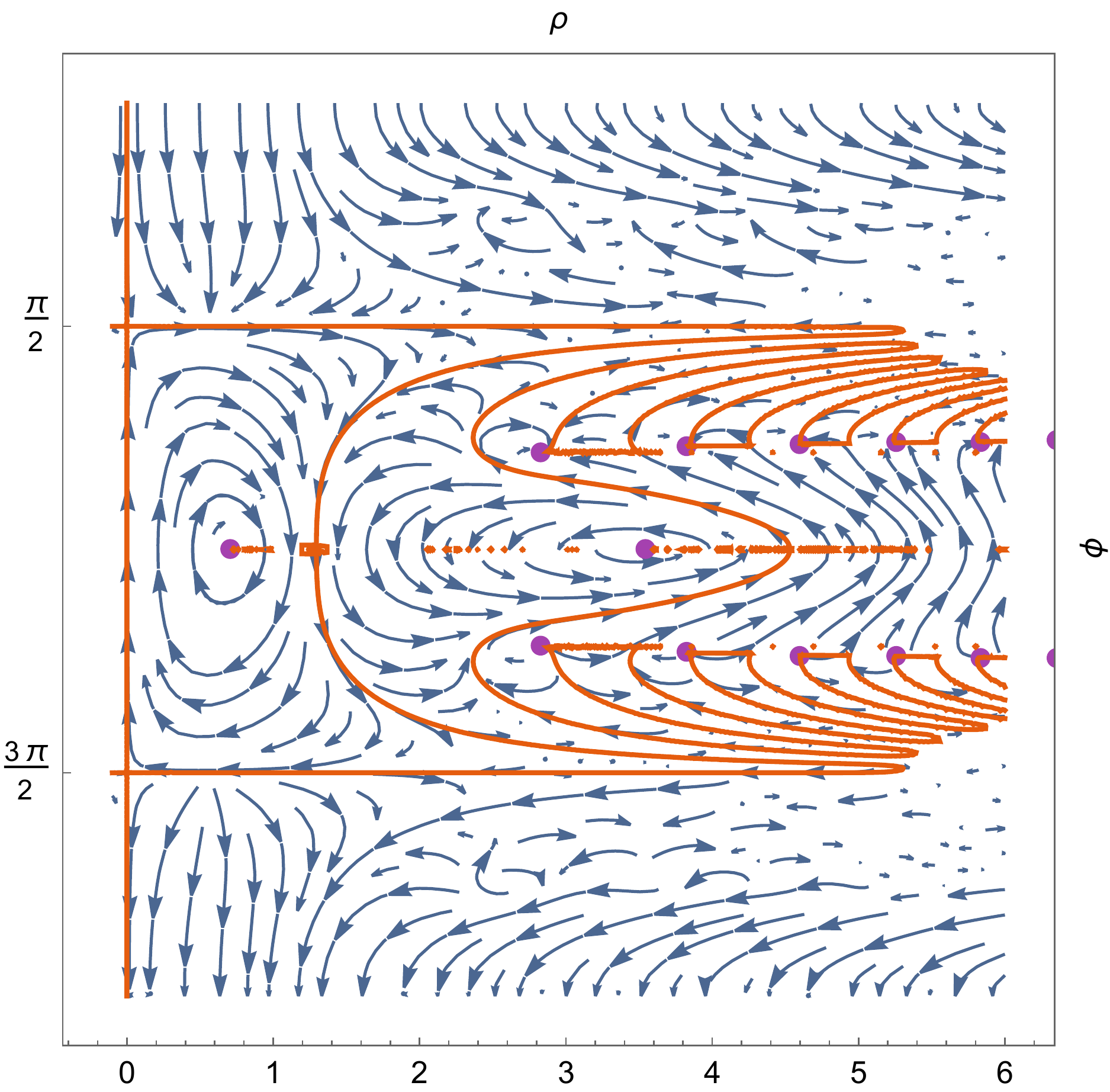}
         \caption{Polar coordinates}
         \label{fig:stokes-diag-nc-ex}
     \end{subfigure}
      \captionsetup{width=.9\textwidth}
       \caption{Stokes diagrams for the Eq.\ \eqref{eq:NC-BH-shape} BH;
    $\star\dif\Re[F(x, y)]$--blue vector fields
    with $P=1/2$. }
        \label{fig:stokes-diag-nc-full}
\end{figure}

The complex extended horizons can be calculated numerically using $f(r)=0$. 
The distribution of complex horizons depends on the value of $\abs{r}$,
which can be illustrated by the argument principle\footnote{The difference between numbers of zeros and 
poles of a meromorphic function $w=f(z)$ equals winding index of mapping curve $f\circ\gamma$ around $w=0$ \cite{shabat2004intro}.}, see Fig.\ \ref{fig:winding}, 
where we show winding indices for the images of circles with different moduli $\abs{r}$. 
The number of complex horizons increases with  increasing modulus  $\abs{r}$. 
Moreover,  we can find a property of the horizons,
that is, if $r$ is a horizon, 
 its conjugate $r^*$ is also a horizon because $\gamma^*(3/2, z^2)=\gamma(3/2, (z^*)^2)$, and thus, $0=f^*(r)=f(r^*)$, 
see Fig.\ \ref{fig:stokes-diag-nc}.

As the modulus of $r=x+\mi y$ becomes large,
the complex horizons can be approximately computed  by solving the system
\begin{equation}
4 \me^{y^2-x^2} \cos (2 x y)+1=0,\qquad
\sin (2 x y)=0
\end{equation}
which gives
\begin{equation}
2 x y=2 \pp n+\pp,\qquad 
x^2-y^2=2\ln (2)
\end{equation}
with $n\in\mathbb{Z}$.
We can also verify that there are neither critical nor semi-critical points in this model,
because $\gamma\left(3/2,  r^2\right)/r$ is an entire function.
In particular, $1/f$ approaches $1$ as $r\to 0$.
Therefore, the monodromy of the asymptotic solutions along any closed curve is trivial, 
which leads to a similar result to that of the above example, $1=\exp(\omega/T^+_{\rm H})$.

 However, if we consider the small-charge approximation $P\sim 0$, i.e., if we treat the RBH model as a singular one,
we obtain
 \begin{equation}
 \label{eq:small-charge}
 f(r)\sim 1-\frac{\sqrt{\pp }}{P r}
 +\frac{2 }{P}\me^{-r^2},
 \end{equation}
which has a critical point at zero and one real horizon.
In other words, the Stokes diagram of this model in the small-charge approximation 
is similar that of the Schwarzschild BH.
The  tortoise coordinate and effective potential close to zero are, respectively, 
\begin{equation}
z\sim -\frac{r^2}{4 \sqrt{\pp }},\qquad
V\sim -\frac{\pp }{P^2 r^4}.
\end{equation}
The master equation becomes
\begin{equation}
\left(  \frac{\dif{}^2 }{\dif z^2}+\omega^2
-\frac{1}{16 P^2 z^2}\right)\psi( z)=0,
\end{equation}
and the monodromy relation is then 
\begin{equation}
-\left[1+2 \cos (\pp  \nu )\right]=\me^{\omega /T_{\rm H}}
\end{equation}
with 
\begin{equation}
\nu = \frac{1}{2} \sqrt{\frac{1}{P^2}+4},\qquad
T_{\rm H} =\frac{\sqrt{\pp } \me^{r_{\rm H}^2}-4 r_{\rm H}^3}{4 \pp ^{3/2} \me^{r_{\rm H}^2} r_{\rm H}-8 \pp  r_{\rm H}^2}.
\end{equation}

As we find in this section, the analytical spectrum of the AQNMs reduces to an extreme simple form because the monodromy of the asymptotic solutions along any closed Stokes lines is trivial.

\section{Relationship between the monodromy method and WKB approach}
\label{sec:comparison}

\begin{figure}[!ht]
     \centering
     \begin{subfigure}[b]{0.455\textwidth}
         \centering
         \includegraphics[width=\textwidth]{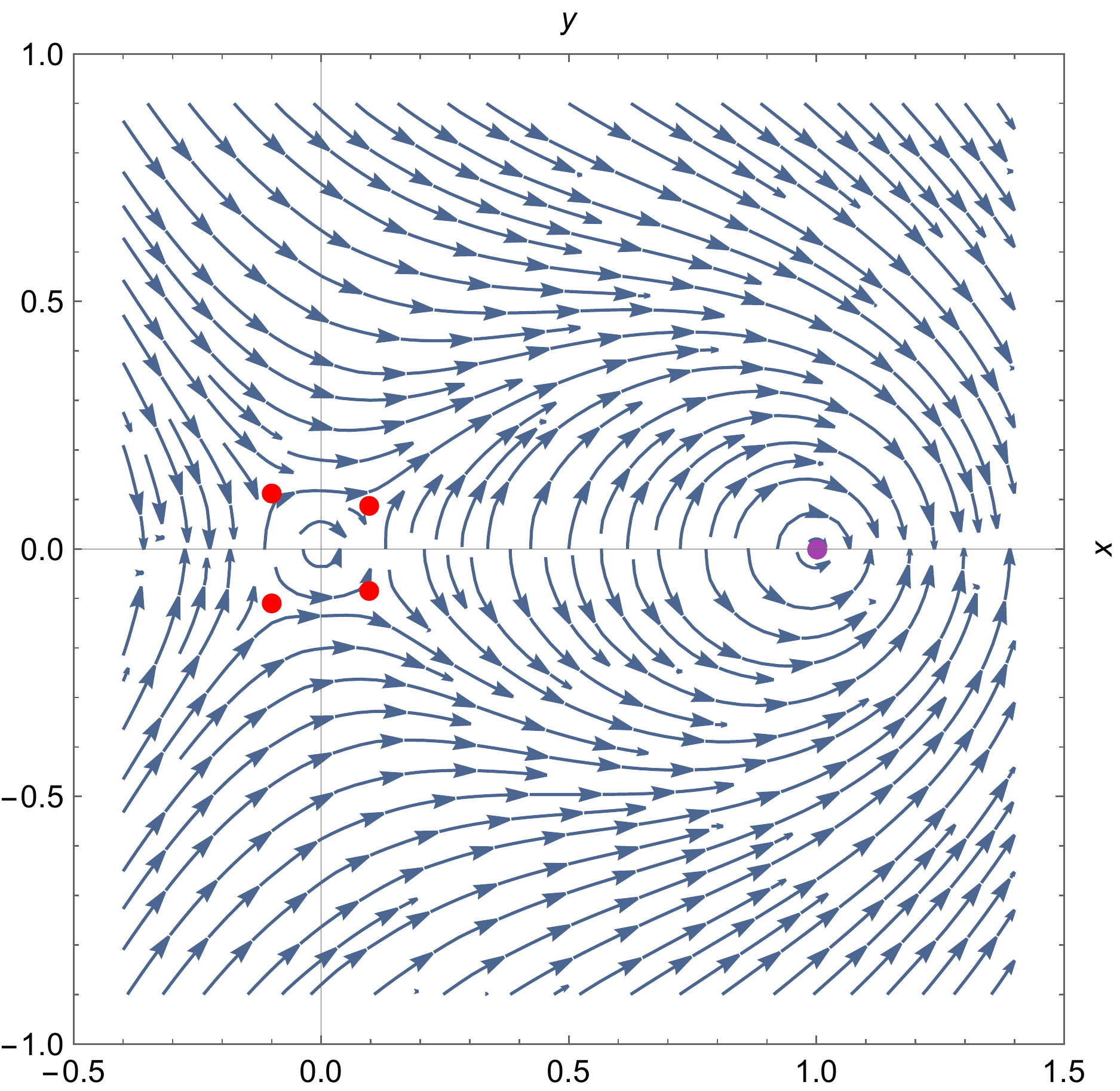}
         \caption{Schwarzschild BH}
         \label{fig:stokes-diag-ah-sch}
     \end{subfigure}
     \begin{subfigure}[b]{0.445\textwidth}
         \centering
         \includegraphics[width=\textwidth]{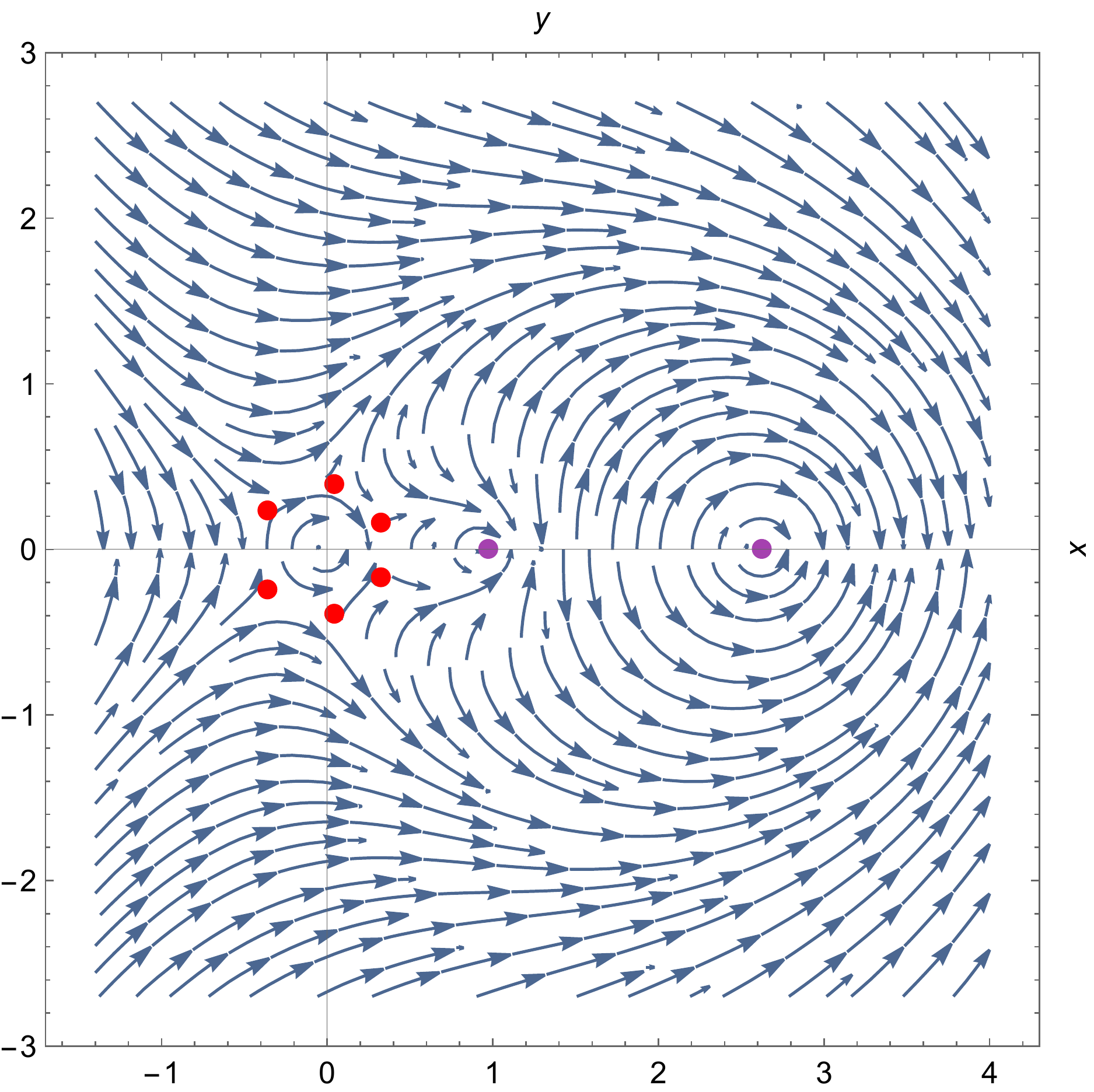}
         \caption{Reissner--Nordstr\"om BH}
         \label{fig:stokes-diag-ah-rn}
     \end{subfigure}
      \captionsetup{width=.9\textwidth}
       \caption{Stokes field $\{\Re[Q],-\Im[Q]\}$. The purple points denote horizons, whereas the red points are zeros of $Q$.}
        \label{fig:stokes-diag-ah}
\end{figure}

To study the AQNMs using the {\em complex} WKB approach, 
Andersson and Howls used another method to diagonalize the differential operator in the master equation, 
where the radial coordinate, unlike the tortoise coordinate, is no longer mutivalued \cite{Andersson:2003fh}. 
The diagonalized master equation in Andersson and Howls' approach is
\begin{equation}
\frac{\dif^2\Phi}{\dif r^2} +Q^2 \Phi =0
\end{equation}
with 
\begin{equation}
Q^2 = f^{-2}  Q^2_0,\qquad
Q^2_0=\left[
\omega^2-V(r)
+\frac{1}{4} \left(\frac{\dif f}{\dif r}\right)^2
-\frac{1}{2} f\left(\frac{\dif^2 f}{\dif r^2}\right)
\right]
\end{equation}
where $V(r)$ is the same as in Eq.\ \eqref{eq:potential}, 
and $\Phi$ connects with $\psi$ via $\Phi=f^{1/2}\psi$.
The Stokes lines are defined by 
\footnote{We use opposite convention for the definitions of Stokes and Anti-Stokes lines.}
\begin{equation}
\label{eq:stokes-ah}
\Im\left[
\int^{r}_{r_0} Q(r') \dif r'
\right]=0.
\end{equation}
Generally, the integral in Eq.\ \eqref{eq:stokes-ah} cannot be calculated analytically for RBHs. 
Therefore, we use the {\em WKB Stokes field} $\{\Re[Q],-\Im[Q]\}$ 
 to depict the Stokes portrait, see Fig.\ \ref{fig:stokes-diag-ah}, 
 where the Stokes field for Schwarzschild and RN BHs are shown.

We use the term \enquote{WKB Stokes field} to distinguish it from
the one we discuss in the monodromy method. 
However, for a given BH model, 
these two fields are related.
To see it clearly, let us take the damping limit $\omega\to-\mi \infty$
in the WKB Stokes field, 
which gives us $Q\sim -\mi 
\abs{\Im(\omega)}/f(r)$ for $r\ge r_{\rm H}$.
Thus, the  WKB Stokes field in the damping limit becomes
\begin{equation}
\{\Re[Q],-\Im[Q]\}\sim\abs{\Im(\omega)} \left\{  \Im\left[1/f(r)\right],  \Re\left[1/f(r)\right]\right\}
\end{equation}
which is the scaled Stokes field in the monodromy method, Eq.\ \eqref{eq:stokes-field}.
Furthermore, for singular BHs, the critical points (zeros) of $Q$  converge to the origin (essential singularities) of BHs
as $\Im(\omega)\to \infty$,
and the WKB Stokes field becomes the Stokes field in the monodromy method.

\begin{figure}[!ht]
     \centering
     \begin{subfigure}[b]{0.32\textwidth}
         \centering
         \includegraphics[width=\textwidth]{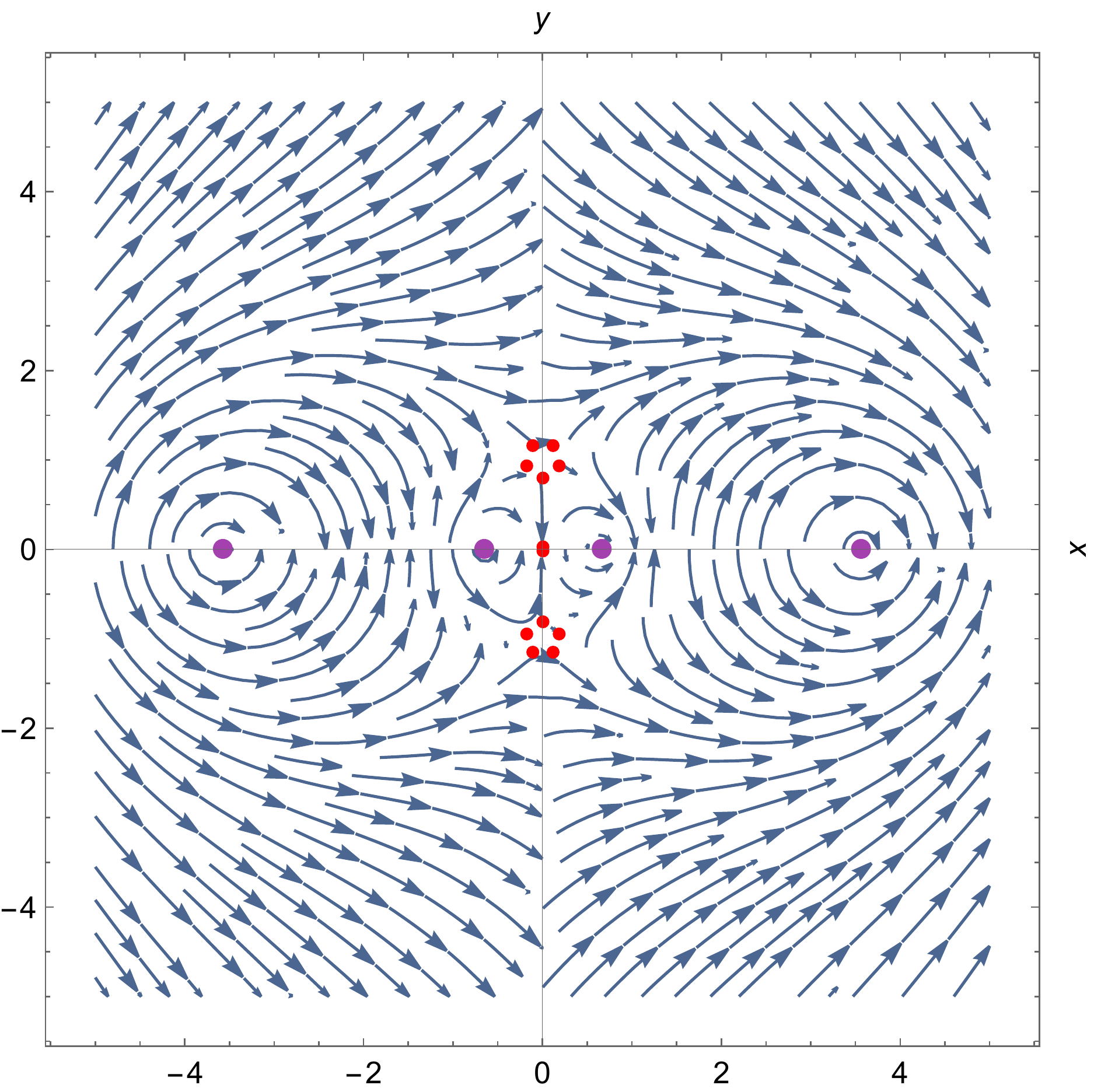}
         \caption{Bardeen BH}
         \label{fig:stokes-diag-ah-bardeen}
     \end{subfigure}
     \begin{subfigure}[b]{0.32\textwidth}
         \centering
         \includegraphics[width=\textwidth]{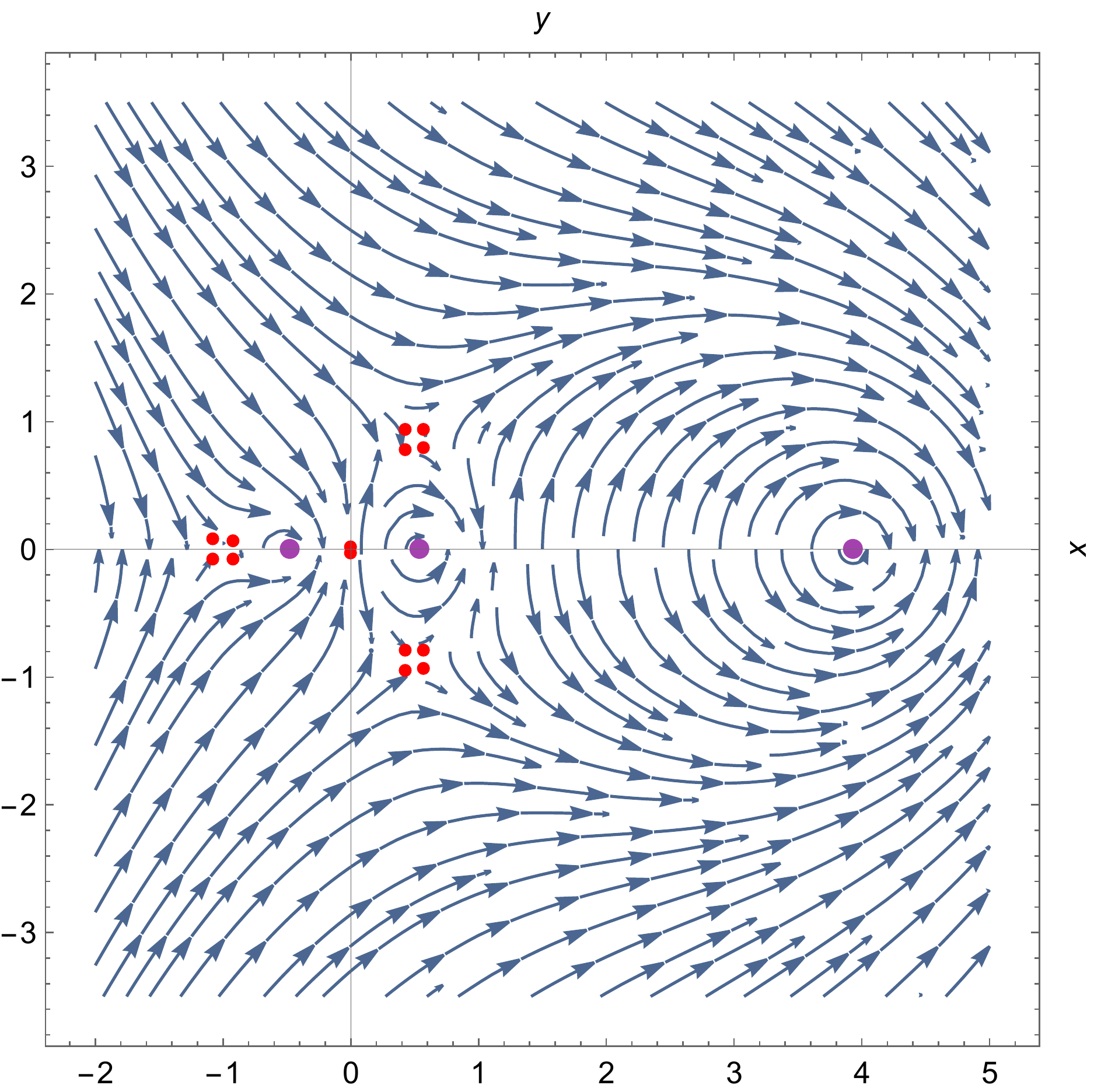}
         \caption{Hayward BH}
         \label{fig:stokes-diag-ah-hayward}
     \end{subfigure}
     \begin{subfigure}[b]{0.32\textwidth}
         \centering
         \includegraphics[width=\textwidth]{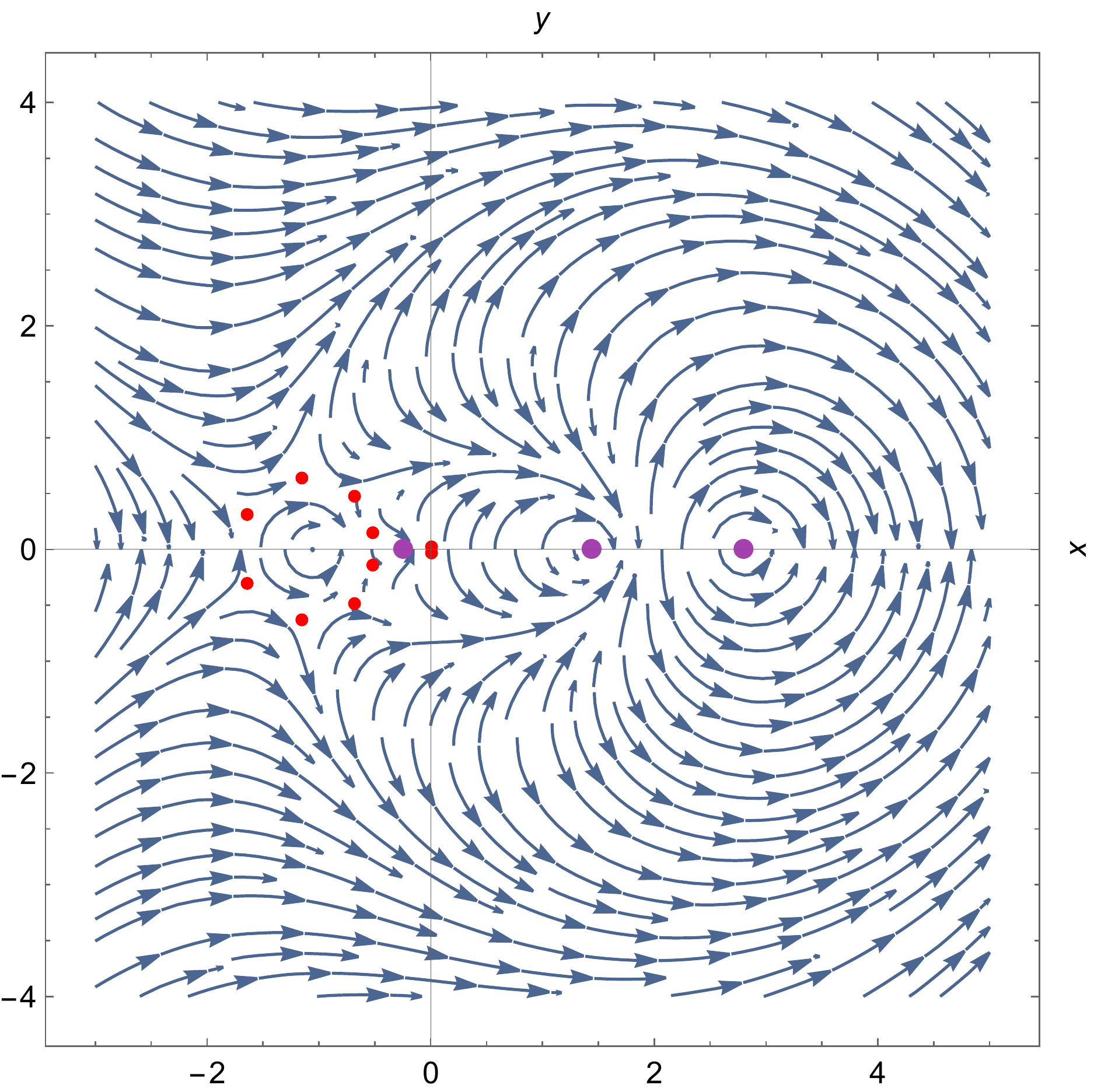}
         \caption{Third class BH}
         \label{fig:stokes-diag-ah-thrid}
     \end{subfigure}
      \captionsetup{width=.9\textwidth}
       \caption{Stokes field $\{\Re[Q],-\Im[Q]\}$. The purple points denote horizons, whereas the red points are zeros of $Q$.}
        \label{fig:stokes-ned-ah}
\end{figure}

The situation for RBHs is slightly different from the above case for singular BHs.
$Q_0^2$ of RBHs has two zeros around the origin $r=0$ owing to $V\sim r^{-2}$ (see Fig.\ \ref{fig:stokes-ned-ah}),
 even though $r=0$ is not a zero of $f^{-2}$.
 However, these two zeros do not converge to the origin
 but disappear as $\omega \to -\mi \infty$
 because $\omega$ dominates in $Q_0^2$ at that moment.
 As a result, $r=0$ becomes a regular point in the damping limit. In other words, the AQNMs via the complex WKB approach, like in the monodromy method, are not dependent on the behavior around the BH center.

\section{Conclusions and outlook}
\label{sec:concl}

In this study, to calculate the AQNMs of RBHs, we apply the Stokes field to extensively investigate the local characteristics of the Stokes lines and classify RBHs based on the types of their complex singularities, which
arise as a result of the analytical continuation of the radial coordinate into the complex plane.

From the calculation, we find several novel aspects of the AQNMs of RBHs, i.e., the analytical forms of the asymptotic frequency spectrum are not universal for spherically symmetric RBHs with single shape functions 
and do not depend on the multipole number $l$ because
$r=0$ is not a point on the Stokes lines;
even if $r=0$ is a point on the Stokes lines, it must not be an
accumulation point, such that the rotation of the asymptotic solution around $r=0$ is trivial and $l$ does not appear in the formula of the AQNM. 

The forms of asymptotic frequency may depend on the structures of the Stokes portraits.
\begin{enumerate}
\item The existence of singularities.
The absence of singularities leads to a trivial monodromy, 
such that the asymptotic frequency is purely imaginary.
\item The rotation angle of the asymptotic solutions around singularities. If the rotation angle is trivial, the asymptotic frequency also has only the imaginary part.
\item The trajectory of the asymptotic solutions along the Stokes lines. It is not the topology of the Stokes line that plays a decisive role, but the way of bypassing the trajectory.
\end{enumerate}

In a broader sense,
specific research on AQNMs relates to several mathematical topics, such as
transcendental curves and the value distribution of holomorphic functions. As shown in this study, even the simplest Stokes lines obtained from the Schwarzschild BH cannot be depicted by a polynomial, whereas the Stokes lines for RBHs are usually not integrable and have no analytical expressions. 

On the one hand, to provide aspects of the Stokes lines, we attempt several numerical methods in this study to remove the integral of the tortoise coordinate, e.g., Newton–Cotes quadrature rules. Nevertheless, the results obtained from these numerical methods more or less lose some important information on the Stokes lines.

On the other hand, the situation of the complex singularities (curvature and coordinate singularities) of RBHs becomes intricate because the Stokes lines must have self-intersections at curvature singularities, and their closed parts must surround the coordinate singularities. 
Thus, the distribution of the zeros and poles of the shape functions directly affect the appearance of the Stokes lines. Clarification of the value distribution may help us construct information on the Stokes lines. Therefore, we will devote 
ourselves to developing more effective approaches to obtaning the feature of Stokes lines in future studies.

Finally, because our main motivation for studying the AQNMs is to obtain the quantum entropy spectrum of RBHs, our subsequent work will focus on how to derive the correct quantum entropy spectrum based on the AQNMs obtained in this study.

\section*{Acknowledgments}

C.L.\ acknowledges support from the National Natural Science Foundation of China under
Grant No.\ 12175108.

\appendix


\bibliographystyle{utphys}

\bibliography{references}

\end{document}